\documentclass[onecolumn,prx,superscriptaddress,showpacs,12pt]{revtex4}
\usepackage{times}
\usepackage{amsmath}
\usepackage{amssymb}
\usepackage{graphicx}
\usepackage{subfigure}
\usepackage{verbatim}
\usepackage{amsfonts}
\usepackage{dcolumn}
\usepackage{bm}
\usepackage{color}
\usepackage[colorlinks,citecolor=blue]{hyperref}

\usepackage{mathrsfs}
\usepackage{setspace}
\usepackage{adjustbox}

\newcommand{\beq}{\begin{eqnarray} }
\newcommand{\eeq}{\end{eqnarray} }
\newcommand{\Beq}{\begin{eqnarray*} }
\newcommand{\Eeq}{\end{eqnarray*} }
\newcommand{\Bmat}{\left(\begin{matrix}}
\newcommand{\Emat}{\end{matrix}\right)}
\newcommand{\up}{\uparrow}
\newcommand{\dn}{\downarrow}

\newcommand{\R}{ \color{red}}
\newcommand{\upcite}[1]{\textsuperscript{\textsuperscript{\cite{#1}}}}
\usepackage{titlesec}

\begin{document}
\begin{center}

\textbf{{\large Evidence for field induced quantum spin liquid behavior in a
spin-1/2 honeycomb magnet}}
\end{center}

\begin{center}
Gaoting Lin,$^{1,11}$ Mingfang Shu,$^{1,10,11}$ Qirong Zhao,$^{2,11}$ Gang
Li,$^{3}$ Yinina Ma,$^{3}$ Jinlong Jiao,$^{1}$ Yuting Li,$^{1}$ Guijing
Duan,$^{2}$ Qing Huang,$^{4}$ Jieming Sheng,$^{5}$ Alexander I. Kolesnikov,$^{6}$
Lu Li,$^{7}$ Liusuo Wu,$^{5}$ Hongwei Chen,$^{8}$ Rong Yu,$^{2,9}$ Xiaoqun
Wang,$^{1}$ Zhengxin Liu,$^{2,9}$* Haidong Zhou,$^{4}$* Jie Ma$^{1,9}$*.
\end{center}

$^{1}$Key Laboratory of Artificial Structures and Quantum Control, School of
Physics and Astronomy, Shanghai Jiao Tong University, Shanghai 200240, China.

$^{2}$Department of Physics, Renmin University of China, Beijing 100872, China.

$^{3}$Beijing National Laboratory for Condensed Matter Physics, Institute of
Physics, Chinese Academy of Sciences, Beijing 100190, China.

$^{4}$Department of Physics and Astronomy, University of Tennessee, Knoxville,
Tennessee 37996, USA.

$^{5}$Shenzhen Institute for Quantum Science and Engineering (SIQSE) and
Department of Physics, Southern University of Science and Technology (SUSTech),
Shenzhen, Guangdong 518055, China.

$^{6}$Neutron Scattering Division, Oak Ridge National Laboratory, Oak Ridge, TN
37831, USA.

$^{7}$Department of Physics, University of Michigan, Ann Arbor, Michigan 48109,
USA.

$^{8}$College of Materials Science and Chemistry, China Jiliang University,
Hangzhou

310018, China

$^{9}$Collaborative Innovation Center of Advanced Microstructures, Nanjing
University, Nanjing 210093, Jiangsu, China

$^{10}$Guangdong Provincial Key Laboratory of Extreme Conditions, Dongguan,
523803, China

$^{11}$These authors contributed equally

{\raggedright
*Correspondence:  liuzxphys@ruc.edu.cn(Z.L.),
hzhou10@utk.edu(H.Z),jma3@sjtu.edu.cn(J.M.)
}
\pagebreak{}

\begin{center}
\textbf{ABSTRACT}
\end{center}

One of the most important issues in modern condensed matter physics is the
realization of fractionalized excitations, such as the Majorana excitations in
the Kitaev quantum spin liquid. To this aim, the 3\textit{d}-based Kitaev
material Na$_{2}$Co$_{2}$TeO$_{6}$ is a promising candidate whose magnetic phase
diagram of \textbf{B} // \textbf{a*} contains a field-induced intermediate
magnetically disordered phase within 7.5 T $<$ $\vert{}$B$\vert{}$ $<$ 10 T. The
experimental observations, including the restoration of the crystalline point
group symmetry in the angle-dependent torque and the coexisting magnon
excitations and spinon-continuum in the inelastic neutron scattering spectrum,
provide strong evidence that this disordered phase is a field induced quantum
spin liquid with partially polarized spins. Our variational Monte Carlo
simulation with the effective K-J$_{1}$-$\Gamma{}$-$\Gamma{}$'-J$_{3}$ model
reproduces the experimental data and further supports this conclusion.

{\raggedright
\textbf{Keywords: }Kitaev quantum spin liquid, bond-dependent frustrated magnet,
quantum fluctuations under magnetic fields, honeycomb lattice.
}
\pagebreak{}

\textbf{INTRODUCTION }
\vspace{8pt}

    Quantum spin liquids (QSLs) are exotic phases of matter resulting from competing
interactions or geometric frustration. Due to the long-range quantum entanglement
in the QSL ground states, interesting phenomena can arise, such as the collective
excitations with fractional quantum numbers and the emergence of gauge
fluctuations.$^{1,2 }$A lot of efforts have been made to search for QSL phase
with Heisenberg-type exchange interactions in geometrically-frustrated systems,
including the triangular lattice, kagom\'{e} lattice, and three-dimensional
structures such as the hyper-kagom\'{e} lattices.$^{2-8}$ Meanwhile, anisotropic
interactions resultant from spin-orbital coupling (SOC) have attracted more and
more interests.$^{1,9-32}$ A paradigmatic example is the exactly solvable Kitaev
model on the honeycomb lattice which hosts QSL ground state and
Majorana-fermion-like elementary excitations.$^{8}$

Recently, exciting progress has been made on the Kitaev spin liquid candidates
from the 3\textit{d}/4\textit{d}/5\textit{d} transition metal Co/Ru/Ir for the
low-energy effective interactions, which contain the Kitaev-type exchange terms
due to the peculiar lattice structure and the SOC.$^{1,15-31,34-38}$ However,
owing to the existence of non-Kitaev interactions, all of these materials exhibit
zigzag antiferromagnetic (AFM) order at low temperatures. Besides the Heisenberg
exchanges, the off-diagonal symmetric interactions of the \textit{$\Gamma{}$} and
\textit{$\Gamma{}$}${'}$ terms were proposed to construct their low-energy
effective model.$^{ 16-19,23,24,36}$ Unlike the Ru- and Ir-
materials,$^{19,36,38}$ the 3\textit{d} orbitals in the Co-based materials are
more compact and the contributions from the SOC channels
\textit{t}$_{2\textit{g}}$-\textit{e$_{g}$} and
\textit{e$_{g}$}-\textit{e$_{g}$}$_{ }$can weaken the \textit{$\Gamma{}$} and
\textit{$\Gamma{}$}${'}$ exchanges.$^{24,38}$

One of the most representative Kitaev materials among the 3\textit{d}-cobalt
magnets is the Na$_{2}$Co$_{2}$TeO$_{6}$ (NCTO),$^{ 1,24,29,34,35,39-44}$ in
which the honeycomb layers are formed by the magnetic Co$^{2+}$ ions surrounded
by the O$^{2-}$ octahedrons, Figures 1A-B. The principal reciprocal vectors
\textbf{a*} (crystallographic vector \textbf{a}) direction is parallel
(perpendicular) to the Co-Co bond, which corresponds to the [$\bar{1}$10]
([$\bar{1}$2$\bar{1}$]) direction in the spin coordinate, Supplementary Figure
S1. NCTO presents a zigzag AFM order below \textit{T}$_{N}$ $\approx{}$ 26 K and
another two anomalies at \textit{T}$_{F}$ $\approx{}$ 15 K and \textit{T}*
$\approx{}$ 7 K, Figures S2A-B.$^{1,35,41-44}$ The zigzag order in NCTO can be
easily suppressed by a magnetic field parallel to the \textbf{a*}-axis, leading
to an intermediate field-induced magnetically disordered phase above
\textasciitilde{} 7.5 T before entering a trivial polarized phase near 10
T.$^{1,44}$ The exact nature of this intermediate phase, most intriguingly,
whether it belongs to a QSL, is still illusive and deserves further
investigation.$^{1}$

The study of zero-field spin-wave excitations of NCTO indicates that while the
AFM third-neighbor Heisenberg exchange interaction \textit{J}$_{3}$ is fairly
large,$^{1,29,33}$ the Kitaev term cannot be ignored.$^{24,29}$ Several
theoretical works have estimated the value of \textit{K}, however, it varies from
large to small, even its sign from ferromagnetic to
AFM.$^{1,24,29,34,35,38,41,45,46}$ It is reasonable to expect that the SOC caused
bond-dependent interactions, including the Kitaev term, play an important role in
understanding the rich phase diagram of NCTO in magnetic fields, Figure1C.

In the present work, we studied the nature of the field-induced intermediate
magnetically disordered phase of a single-crystal NCTO via magnetic torque and
inelastic neutron scattering (INS) spectroscopy. Under low temperatures and low
fields, the torque is very weak and exhibits a 2-fold (C$_{2}$) symmetric angular
dependence, which confirms the AFM long-range order. The AFM order vanishes above
7.5 T as the lattice 6-fold (C$_{6}$) symmetry is restored in the angular
dependence of the torque, which is verified by the disappearance of Bragg peaks
at the M-point at B = 8 T, Figures 1D-E. The material enters the polarized phase
at 10 T where a phase transition is observed in the differential magnetic
susceptibility as well as the differential torque, Figure 1G. The region between
7.5 T and 10 T is shown to be a field-induced QSL phase with partial spin
polarization and strong quantum fluctuations. With an 8 T magnetic field along
the \textbf{a*} direction, the intensity of INS spectra at the M-point is
suppressed, while gapped spin-wave bands show up at 1.5 meV\textasciitilde{}2.5
meV and 3 meV\textasciitilde{}4 meV in the vicinity of the $\Gamma{}$-point
(resulting from the partial polarization of the spins) and an intense
`$\Lambda{}$' shape spinon continuum appears at 4 meV\textasciitilde{}8 meV.
These features are consistent with a theoretically computed dynamical structure
factor of a field-induced partially polarized QSL phase.

\vspace{8pt}
\textbf{MATERIALS AND METHODS}
\vspace{8pt}

\textit{Sample preparation and characterization}. The high-quality single
crystals were grown by the flux method. The polycrystalline sample of NCTO was
mixed with the flux of Na$_{2}$O and TeO$_{2}$ in molar ratio of 1:0.5:2 and
gradually heated to 900 $^\circ{}$C at 3 $^\circ{}$C/min in air after grinding.
The sample was retained at 900 $^\circ{}$C for 30 h, and was cooled to a
temperature of 500 $^\circ{}$C at the rate of 3 $^\circ{}$C/h. The furnace was
then shut down$^{1}$.

\textit{Magnetization and heat capacity}. The magnetization measurements were
performed by using a vibrating sample magnetometer (VSM) in the physical
properties measurement system (PPMS Dynacool-9 system, Quantum Design) with field
up to 9 T. The heat capacity measurements were carried out using the relaxation
method in another PPMS with field up to 13 T. The magnetization and heat capacity
could be found in the supplementary Figure S2.

\textit{Magnetic torque}. The magnetic torque measurements were carried out
using piezo-resistive sensor made by Quantum Design, external bridge excitation
and Lock-in amplifier readout were utilized. An oriented NCTO single crystal was
mounted onto the sensor. The magnetic field was applied in the ab plane, as
illustrated in Supplementary Figure S3. Both angular and magnetic field dependent
torque measurements were carried out. The low temperature and magnetic field
environment were provided by either a Quantum Design PPMS-9 or a top-loading
18T-320 mK system.

\textit{Inelastic neutron scattering}. INS experiments were performed using the
SEQUOIA time-of-flight spectrometer at the Spallation Neutron Source, Oak Ridge
National Laboratory, USA.$^{ 47,48}$ About 0.559 g samples were fixed on an
aluminum sheet with 3 $\times{}$ 6 $\times{}$ 0.05 cm$^{3}$ in size, and
co-aligned in the (HHL) scattering plane with \textbf{B} // \textbf{a*},
Supplementary Figure S8. The sample was inserted in a liquid-helium cryostat,
reaching a base temperature of T = 2 K. Measurements at 2 K with applied magnetic
fields B = 0 T and 8 T were performed by rotating the sample in steps of
1$^\circ{}$ with E$_{i}$ = 18 meV and choppers in high-resolution mode, yielding
a full-width at half-maximum (FWHM) elastic energy resolution of about 0.41 meV.
When the magnet was removed, we collected again the INS data at 0 T and 4.9 K,
which also were performed by rotating the sample in steps of 1$^\circ{}$ with
E$_{i}$ = 18 meV and choppers in high-resolution mode. In order to subtract the
background, the INS data were collected at 90 K with or without magnet.

\textit{Variational Monte Carlo simulation}. The VMC method is a variational
approach using Gutzwiller projected mean field states as trial wave functions of
spin models. The mean field state is obtained in the slave particle
representation, where the spin operators are represented in bilinear form of
fermions under a particle number constraint. The mean field parameters are not
obtained self-consistently, but are treated as variational parameters whose
optimal values are determined by minimizing the trial energy. The trial energy
and physical quantities (including the correlation functions) of the Gutzwiller
projected state are obtained using Monte Carlo simulations.

\begin{figure*}[htbp]
\small
\includegraphics[width=.9\linewidth]{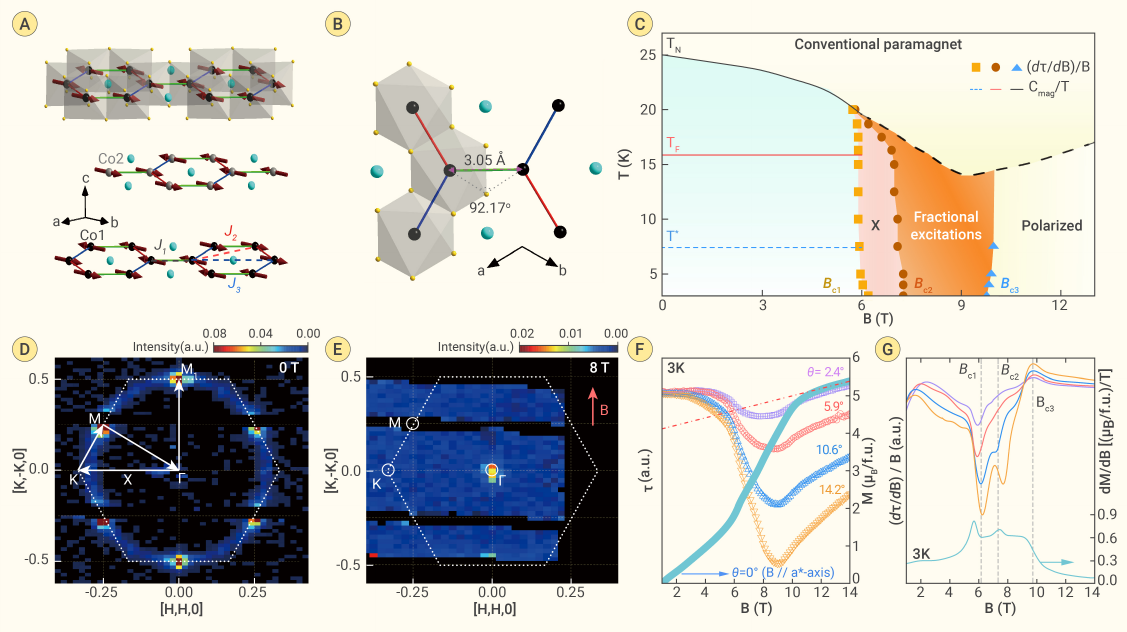}
\caption{$\textbf{Structure, magnetic properties and
temperature-field phase diagram of NCTO.}$ (A) Three-dimensional stacking of the Co
honeycomb layers. Honeycomb network shows the Co-Co bonds (red/blue/green) and
edge-shared CoO$_{6}$ octahedra (the black, grey, cyan and golden spheres
represent Co1, Co2, Te and O atoms, respectively). The Co1 and Co2 honeycomb
layers present the ABAB-type layer stacking along the \textit{c} axis. The dark
red refers to the direction of moments that are in the \textit{ab} plane and
parallel to \textit{b}-axis, indicating a zigzag AFM ground state.$^{42,43}$ (B)
A honeycomb network with three selected adjacent edge-shared CoO$_{6}$ octahedra.
In the P6$_{3}$22 structure, Co ions form a perfect honeycomb lattice with an
equal 92.17$^\circ{}$ Co-O-Co bond angle and the nearest-neighbor Co-Co bond
length d$_{Co-Co}$ = 3.05 \AA{}. (C) Temperature-field phase diagram with
\textbf{B} parallel to Co-Co bonds. The phase boundaries are deduced from the
temperature-dependent magnetic specific heat \textit{C}$_{mag}$/\textit{T} with
\textbf{B} // \textbf{a*} (Supplementary Figure S2E) and field-dependent
differential magnetic torque $\frac{1}{B}\frac{d\tau{}}{dB}$ with 10.6$^\circ{}$
away from \textbf{a*} (Supplementary Figure S7). When the AFM order is suppressed
by field, the sharp discontinuity in \textit{C}$_{mag}$/\textit{T} at
\textit{T}$_{N}$ labeled by the black solid line becomes crossover type wide
peaks presented by the dashed dark line. The critical fields \textit{B}$_{C1}$,
\textit{B}$_{C2}$, and \textit{B}$_{C3}$ are determined by the
$\frac{1}{B}\frac{d\tau{}}{dB}$ curves at 10.6 degree. Elastic neutron scattering
results integrated over L = [-2.5, 2.5] and E = 0 = $\pm{}$ 0.05 meV at 0 T (D)
and 8 T (E). The white dashed lines represent the Brillouin zone boundaries. The
high symmetry points $\Gamma{}$, X, K, M and M$_{1}$ are marked in (D). The red
arrow in (E) shows the applied magnetic field \textbf{B} // \textbf{a*}-axis. (F)
Field dependence of magnetic torque of NCTO measured at 3 K with field along
selected angles. The solid lines are the polynomial fitting of magnetic torque
ratio. (G) The detivative $\frac{1}{B}\frac{d\tau{}}{dB}$ curves are calculated
from the fitted data. \textit{$\theta{}$} is the angle between \textbf{B} and the
\textbf{a${_\ast}$} and with respect to the real-space orientation of the Co-Co
bonds. More details see the supplementary Figure S1. For better comparison, the
field dependence of magnetization M(B) (F) and the derivative
\textit{d}M(B)/\textit{d}B (G) are also shown with \textbf{B} // \textbf{a*}. The
red dashed line shows the Van-Vleck paramagnetic background, which suggests that
the saturation field is around B$_{S}$ =12.5 T and the saturation magnetization
is about M$_{S}$ = 2.05$\mu{}$$_{B}$/Co$^{2+}$.}.
\end{figure*}
\setcounter{figure}{1}

\vspace{8pt}
\textbf{RESULTS }
\vspace{8pt}

\textit{Magnetic torque}. The magnetic torque of a sample \textbf{$\tau{}$} =
\textit{$\mathrm{\mu}$}$_{0}$\textit{V}\textbf{M}$\times{}$\textbf{B} is highly
sensitive to the external magnetic field \textbf{B} when the induced
magnetization \textbf{M} is not aligned with \textbf{B}, where
\textit{$\mathrm{\mu}$}$_{0}$ denotes the permeability of the vacuum and
\textit{V} the volume of the sample. Therefore, the torque in a uniform
\textbf{B} is a direct detection of the magnetic anisotropy.

Figures 1F-G respectively show the field dependence of the torque ratio and the
first order derivative for field deviating from \textbf{a*} counterclockwise in
the \textit{ab} plane by angles \textit{$\theta{}$} = 2.4$^\circ{}$,
5.9$^\circ{}$, 10.6$^\circ{}$, 14.2$^\circ{}$. Since
$\frac{1}{B}\frac{d\tau{}}{dB}$ = ${\mu{}}_0V\frac{\partial{}M}{\partial{}B}\times{}n+{\mu{}}_0V\frac{M}{B}\times{}n$,
with \textbf{B} = \textit{B}\textbf{n} where \textbf{n} is the unit vector along
the field direction, the quantity $\frac{1}{B}\frac{d\tau{}}{dB}$ contains the
information of the off-diagonal differential magnetic susceptibility and is thus
helpful for locating the phase boundaries, Figure 1G.

At low temperatures, three phase transitions can be identified by the anomalies
in the field derivative ($\frac{1}{B}\frac{d\tau{}}{dB}$) where
\textit{B}$_{C1}$ is the transition field from the zigzag- phase to an
intermediate region labeled as `\textit{X}', \textit{B}$_{C2}$ is the critical
field from the \textit{X}-region to a magnetically disordered phase, and the
\textit{B}$_{C3}$ is the threshold of the trivial polarized phase.$^{1}$ The
values of the critical field strength \textit{B}$_{C1}$, \textit{B}$_{C2}$ and
\textit{B}$_{C3}$ slightly vary with the angle \textit{$\theta{}$}, but the
features of the three transitions are qualitatively unchanged. The anomalies of
the $\frac{1}{B}\frac{d\tau{}}{dB}$ curves become weak when \textit{$\theta{}$}
approaches 0$^\circ{}$ (but the three critical fields are still consistent with
the differential susceptibility \textit{d}M/\textit{d}B curve at
\textit{$\theta{}$} = 0$^\circ{}$ in Figure 1G). Therefore, for clarity we choose
the critical fields at \textit{$\theta{}$} = 10.6$^\circ{}$ to construct the
temperature-field phase diagram, Figure 1C. The critical fields obtained from
specific heat measurements with \textbf{B} // \textbf{a*}(Figure S2E), are
comparable with the ones obtained by the torque measurements at
\textit{$\theta{}$} = 10.6$^\circ{}$. Notice that the field-dependent magnetic
torque and magnetization both show obvious hysteresis near \textit{B}$_{C1}$,
Figure S5, indicating the transition is first-order. This is further verified by
the clear hysteresis loop in the angular dependence of the torque around 6 T,
Figure S4.

The angular dependence of the torque \textit{$\tau{}$}(\textit{$\theta{}$})
directly reflects the symmetry of the magnetic status.$^{ 49-51}$ Since
\textbf{a* }is the easy axis, the induced magnetization \textbf{M} is parallel to
\textbf{a*} if \textbf{B }// \textbf{a*} (i.e. for \textit{$\theta{}$} = n
$\times{}$ 60$^\circ{}$, n is an integer), see Figure 2B-D. As the space group of
NCTO is P6$_{3}$22 (No.182) whose point group is D$_{6}$,
\textit{$\tau{}$}(\textit{$\theta{}$}) should exhibit a C$_{6}$ symmetry (namely
2$\pi{}$/6 periodicity) if there is no symmetry breaking. As shown in Figure 2A,
\textit{$\tau{}$}(\textit{$\theta{}$}) only shows a C$_{2}$ symmetry for
\textit{B} = 3 T. This indicates a rotation symmetry breaking in NCTO (from
C$_{6}$ to C$_{2}$) which confirms the AFM long-range order in weak magnetic
fields below\textit{ T}$_{N}$. Since the thermal fluctuations tend to melt the
symmetry breaking orders, the symmetry of \textit{$\tau{}$}(\textit{$\theta{}$})
is expected to increase with increasing temperature and eventually reaches the
C$_{6}$ in the paramagnetic state above \textit{T}$_{N}$. However, as shown in
Figures 2E \& H, above \textit{T}$_{N}$, the symmetry is still C$_{2 }$with a
different orientation. A possible reason for these inconsistences is that the
magnetic field is not perfectly lying in the \textit{ab}-plane (the
\textit{c}-direction is not strictly parallel to the rotation axis, Figure S3),
thus the absolute value of the angle between the field and the \textit{c}-axis
oscillates with a 2-fold periodicity.$^{52}$ Since the effective in-plane and
out-of-plane \textit{g}-factors are different, \textit{g$_{ab}$} = 4.13 and
\textit{g$_{c}$} = 2.3,$^{1}$ the oscillation of the field component along the
\textit{c}-direction results in the two-fold periodic pattern in
\textit{$\tau{}$}(\textit{$\theta{}$}).

Meanwhile, strong magnetic field and quantum spin fluctuations can also suppress
the zigzag order and restore the symmetry. As shown in Figures 2B-D \& G, in the
angle-dependent torque data the 6-fold symmetry indeed shows up above
\textit{B}$_{C1}$ = 6 T at low temperatures with coexisting C$_{2 }$symmetry. The
C$_{6}$ symmetry becomes almost perfect when the AFM order is completely
suppressed at \textit{B}$_{C2 }$ = 7.5 T. Above \textit{B}$_{C3}$ = 10 T, the
magnitude of the torque decreases with field strength for a polarized phase with
diminished quantum fluctuations. The most interesting physics falls in the region
between \textit{B}$_{C2}$ and \textit{B}$_{C3}$, a field-induced disordered state
with fairly strong quantum fluctuations which is likely to be a QSL phase. Later
we will provide theoretical and further experimental evidences to verify the QSL
phase. The region \textit{X} between \textit{B}$_{C1}$ and \textit{B}$_{C2}$ is
considered as a phase with coexisting AFM and topological order, Figure S6. It
should be mentioned that the \textit{$\tau{}$}(\textit{$\theta{}$}) pattern of
the QSL region (above \textit{B}$_{C2}$) still does not show a strict C$_{6}$
symmetry with some mild amplitude modulation of 2$\pi{}$/2 period, Figures 2C \&
F. Those 2-period Fourier components are the same as that of the high-temperature
paramagnetic state, Figure S6, hence, this should also be the issue of field
alignment mentioned above.

\begin{figure*}[htbp]
\small
\includegraphics[width=.9\linewidth]{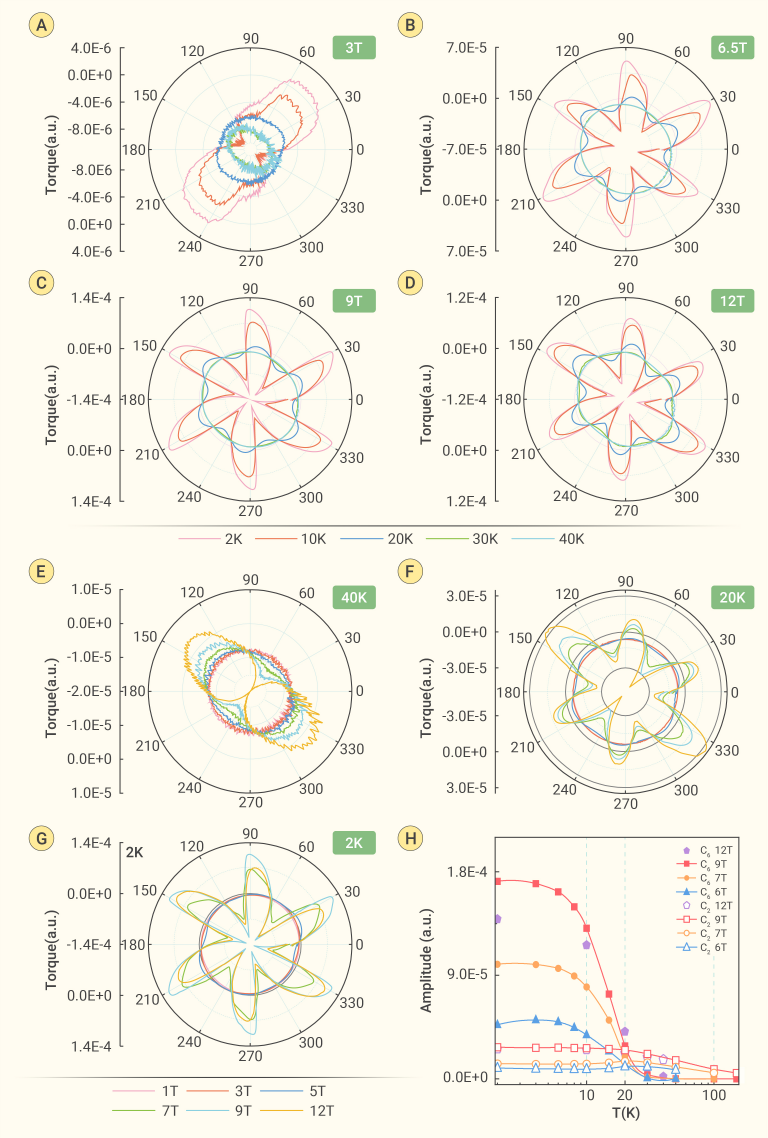}
\caption{$\textbf{Symmetry evolution of magnetic torque with
temperature and magnetic field.}$ (A)-(D) Polar plots of magnetic torque
\textit{$\tau{}$}(\textit{$\theta{}$}) at different temperatures with the fixed
magnetic fields. (E)-(G) Polar plots of magnetic torque
\textit{$\tau{}$}(\textit{$\theta{}$}) at different magnetic fields with the
fixed temperatures. (H) Temperature-dependent amplitude of C6 and C2 symmetry
obtained by Fourier transform of angle-dependent magnetic torque at different
fields.}.
\end{figure*}

This field-induced intermediate QSL phase is supported by the Variational Monte
Carlo (VMC) simulation with a
\textit{K}-\textit{J}$_{1}$-\textit{$\Gamma{}$}-\textit{$\Gamma{}$}${'}$-\textit{J}$_{3}$
model (\textit{J}$_{1}$ is the first-neighbor Heisenberg exchange, the values of
the parameters will be discussed later), where four phases are obtained with
\textbf{B} // \textbf{a*}-axis including the zigzag phase, an intermediated phase
with coexisting magnetic order and topological order, the filed-induced QSL phase
and the polarized trivial phase, Figure S10. Especially, fixing the field's
strength and varying the field's direction in the QSL phase, the induced
magnetization \textbf{M} is parallel to \textbf{B} as \textbf{B} is along the
\textbf{a*}- or \textbf{a}-direction. When the field is deviated from \textbf{a}
or \textbf{a*}, \textbf{M} contains nonzero component in a direction
perpendicular to the field, which gives rise to nonzero magnetic torque. As shown
in Supplementary Figure S12, the simulated magnetization indeed exhibits a 6-fold
periodicity in the QSL region, which is consistent with the experimental data
shown in Figures 2C \& G.

\textit{Neutron scattering}. To further verify the field induced QSL behavior,
we performed scattering measurements at 2 K in the (HHL) plane with \textbf{B} //
\textbf{a*}-direction ([K, -K, 0]) at 0 T and 8 T. As shown in Figure 1D, at zero
field, the magnetic Bragg reflections can be observed at the M-point (such as
[-1/2, 0, 0] and [0, -1/2, 0]), which presents the zigzag AFM order. At 8 T,
these magnetic Bragg reflections at the M-points completely disappear but a new
Bragg reflection appears at the $\Gamma{}$-point for the partial polarization,
Figure 1E.

More interestingly, the applied field also dramatically changes the spin
excitation spectrum. Figure 3A presents the momentum dependence of INS intensity
integrated from 1.5 to 2.5 meV at 4.9 K and 0 T. The ring-shaped spectra are
clearly seen around the M-points, which can be identified as magnon excitations
in the zigzag ordered ground state. At the $\Gamma{}$-point, some excitations
also show up with smaller weight compared to the M-point. Figure 3B shows the INS
intensity integrated from 1.5 to 2.5 meV at \textit{T} = 2 K and
$\vert{}$\textbf{B$\vert{}$ }= 8 T, where the intensity is concentrated at the
$\Gamma{}$-point and represents the edge of a magnon band.

\begin{figure*}[htbp]
\small
\includegraphics[width=.9\linewidth]{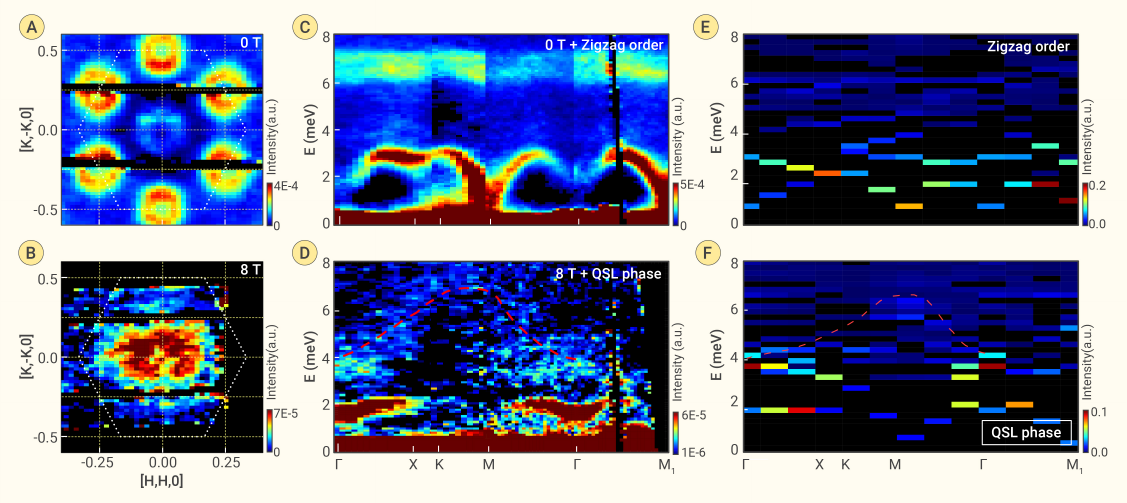}
\caption{$\textbf{Spin-excitation spectra using
fixed incident energy \textit{E$_{i}$} = 18 meV with B $\parallel{}$ a*.}$ Constant-energy scattering at 0 T (A) and 8 T (B), respectively, integrated over
L = [-2.5, 2.5] and E = [1.5, 2.5] meV, projected on the reciprocal honeycomb
plane defined by the perpendicular directions [H, H,0] and [K, -K, 0]. The white
dashed lines represent the Brillouin zone boundaries. (C) and (D) Spin-excitation
spectra along high symmetry momentum directions
$\Gamma{}$-X-K-M-$\Gamma{}$-M$_{1}$ at 5 K for zero field and 2 K for 8 T,
respectively. The color bar indicates scattering intensity with arbitrary unit in
linear scale. The dark red dashed line indicates that an intense `$\Lambda{}$'
shape spinon continuum appears at 4 meV \textasciitilde{} 8 meV in Figure (D).
(E) and (F) Calculated dynamic structure factor for the zigzag AFM order and
field-induced QSL with partially polarized spins, respectively. The black regions
lack detector coverage. The dark red dashed line shows the same `$\Lambda{}$'
shape spinon continuum, which is compared to the experiment at 8T. The data were
collected using the SEQUOIA chopper spectrometer at the Spallation Neutron Source
(SNS).}.
\end{figure*}

To further reveal the intrinsic spin dynamics of the magnetic Co$^{2+}$ ions, we
plot the energy-momentum spectrum of the spin excitations along several
high-symmetry points in the first Brillouin zone, Figure 1D. At zero field, a
gapped magnon band is shown in Figure 3C, where the minimum point of the band has
the largest intensity weight and locates at the M-point as expected. Furthermore,
almost flat magnon bands appear at 6 \textasciitilde{} 7 meV. To understand the
experimental observations, we theoretically study the zero-field dynamic
structure factors in the zigzag ground state and the VMC numerical results are
shown in Figure 3E. The shape of the lowest energy spin-wave band agrees well
with the experimental data, Figure 3C. The band from VMC at 6 meV is almost flat,
and the intermediate weights between the lowest band ($<$ 3 meV) and the flat
band ($>$ 6 meV) show up. These features are consistent with experiment in Figure
3C.$^{34}$

With increasing field, the largest intensity of magnetic excitations shifts from
the M-point to the $\Gamma{}$-point. At 8 T, around the $\Gamma{}$-point, a band
of concave shape shows up at 1.5 meV \textasciitilde{} 2.5 meV and another band
of convex shape appears at 3 \textasciitilde{} 4 meV. These two bands, which look
like the upper lip and the lower lip, are constituted by single-particle-like
magnon excitations. Away from the $\Gamma{}$-point, the weights of the two magnon
bands decay rapidly. Instead, a large piece of continuum is observed at higher
energy, indicating the existence of fractional excitations beyond the
linear-spin-wave theory which predicts only two magnon bands. The continuum
extends to the whole Brillouin zone, and its lower edge is overlapping with the
upper-lip shaped magnon band. Along the $\Gamma{}$-X-K-M-$\Gamma{}$ path, the
bright weights of the continuum in the energy range 4 \textasciitilde{} 8 meV
form a `$\Lambda{}$' shape. The pattern of momentum-energy distribution and the
fairly strong intensity rule out the possibility of two-magnon continuum. Thus,
the continuum is most likely formed by two-spinon fractionalized excitations. The
coexistence of (incomplete) single-particle-like magnon bands and fractionalized
continuum is the most exciting observation of the present work. From the strong
continuum excitations and the C$_{6}$ symmetry of the torque, we infer that NCTO
enters a field-induced QSL phase with partial spin polarization and strong
quantum fluctuations at 8 T and low temperature.

The dynamical structure factor of the field-induced QSL phase obtained from VMC
simulation, Figure3F, captures most of the important features of the neutron
experiment. (1) Both single-particle-like magnon bands and the spinon-continuum
are obtained. The magnon modes are dispersive in-gap two-spinon bound states,
which form two bands. Above the magnon bands a continuum is formed by
fractionalized spinons. The energy ranges of the magnon band and the continuum
agree with experiment. (2) In the vicinity of the $\Gamma{}$-point, the two
magnon bands form the shape of a lower lip and an upper lip. Similar
lip-structure also exists in the linear spin-wave dispersion and is resulting
from the significant \textit{J}$_{3}$ interactions. Nonzero magnon weights appear
at the M- and M$_{1}$-points with energies close to zero. These features agree
with the experiment. (3) The lower edge of the continuum is overlapping with the
upper magnon band. From 4 meV to 8 meV, the bright weights of the continuum form
a shape of `$\Lambda{}$', which qualitatively agrees with the experiment. (4) The
phase has 4-fold topological degenerate ground states on a torus as the emergence
of deconfined Z$_{2}$ gauge fluctuations and the Z$_{2}$ QSL nature of the low
energy physics. The deconfined Z$_{2}$ gauge charges, namely, the
Majorana-fermion like spinons, give rise to the continuum spectrum in the
dynamical structure and interpret the experimental weights at 4 \textasciitilde{}
8 meV. Especially, in the intermediate field region the linear spin wave spectrum
based on a fully polarized state contains imaginary part around the M-points,
Figure S14C, hence, this phase is distinct from the fully polarized phase and is
beyond the description of linear-spin-wave theory. To verify the validity of the
VMC computations, we performed analytic calculations using random phase
approximations (RPA), Figure S15-S16, and the RPA results qualitatively agree
with those of the VMC.

\vspace{8pt}
\textbf{DISCUSSION AND CONCLUSION }
\vspace{8pt}

In our theoretical simulation, we adopted the parameters \textit{J}$_{1}$ =
-1.54 meV, \textit{J}$_{3}$ = 1.32 meV, \textit{K} = 1.408 meV,
\textit{$\Gamma{}$} = -1.32 meV, and \textit{$\Gamma{}$}${'}$ = 0.88 meV, which
are equivalent to \textit{J}$_{1}$ = 0.066 meV, \textit{J}$_{3}$ = 1.32 meV,
\textit{K} = -3.399 meV, \textit{$\Gamma{}$} = 0.286 meV, and
\textit{$\Gamma{}$}${'}$ = 0.077 meV via the dual transformation. This set of
parameters is adopted from the tx+ model,$^{ 46}$ but with enlarged
\textit{J}$_{3 }$and globally multiplied by a constant. The importance of
\textit{J}$_{3}$ has been implied in previous works.$^{1,29,34,35,38-40,46,53}$
In the 3\textit{d} Co-based honeycomb geometry, the hopping integral associated
with the 90$^\circ{}$ \textit{e$_{g}$}-ligand hybridization plays a significant
role through the large $\sigma{}$-type hopping process $t_{Pd}^{\sigma{}}$, which
is particularly relevant for the third neighbor \textit{J}$_{3}$ super-exchanges
in honeycomb materials.$^{24,38}$ Moreover, the ratio $\Delta{}$/$\lambda{}$
between the trigonal crystal field ($\Delta{}$) and the SOC ($\lambda{}$) can
regulate the spin-orbit entanglement.$^{24}$ With the increasing of
$\Delta{}$/$\lambda{}$, the orbital degeneracy is lifted and the spin-orbit
entanglement is suppressed. Powder INS analysis of $\lambda{}$ = 21 meV and
$\Delta{}$ = 13meV with a small ratio $\Delta{}$/$\lambda{}$ $\sim{}$ 0.62
indicates that the spin and orbit are highly entangled.$^{34}$

In summary, based on the magnetic torque and neutron scattering experiments, we
studied the magnetic phase diagram and the nature of each phase of NCTO under
in-plane magnetic fields, especially \textbf{B} $\parallel{}$ \textbf{a*}. At low
temperatures, NCTO belongs to the zigzag AFM phase under field below
\textit{B}$_{C1}$ = 6 T and enters the trivial polarized phase above
\textit{B}$_{C3}$ = 10 T. As the field is between \textit{B}$_{C2}$ = 7.5 T and
\textit{B}$_{C3}$ = 10 T, the restoration of the 6-fold symmetry in the angular
dependence of the torque strongly indicates that NCTO falls in a field-induced
disordered state with strong quantum spin fluctuations. Furthermore, the strong
continuum in the INS spectrum and the magnon-like excitations near the
$\Gamma{}$-point confirm that this disordered state is a QSL state whose spins
are partially polarized. Our theoretically obtained spin excitation spectra from
VMC simulations of the effective
\textit{K}-\textit{J}$_{1}$-\textit{$\Gamma{}$}-\textit{$\Gamma{}$}${'}$-\textit{J}$_{3}$
model, including the dynamical structure factors of the AFM phase and the
partially polarized QSL phase, agree with the experimental data, and support the
field-induced QSL behavior in NCTO. Finally, we identify the intermediate
\textit{X}-region between \textit{B}$_{C1}$ = 6 T and \textit{B}$_{C2}$ = 7.5 T
as a phase with coexisting AFM order and Z$_{2}$ topological order.

\vspace{8pt}
\textbf{REFERENCES}
\vspace{8pt}

1. Lin, G., Jeong, J., Kim, C., et al. (2021). Field-induced quantum spin
disordered state in spin-1/2 honeycomb magnet Na$_{2}$Co$_{2}$TeO$_{6}$. Nat.
Commun. \textbf{12}: 5559. DOI: 10.1038/s41467-021-25567-7.

2. Broholm, C., Cava, R.J., Kivelson, S.A., et al. (2020). Quantum spin liquids.
Science \textbf{367}: 263. DOI: 10.1126/science.aay0668.

3. Shen, Y., Li, Y.-D., Wo, H., et al. (2016). Evidence for a spinon Fermi
surface in a triangular-lattice quantum-spin-liquid candidate. Nature
\textbf{540}: 559-562. DOI: 10.1038/nature20614.

4. Fu, M., Imai, T., Han, T.-H., et al. (2015). Evidence for a gapped
spin-liquid ground state in a kagome Heisenberg antiferromagnet. Science
\textbf{350}: 655-658. DOI:10.1126/science.aab2120.

5. Okamoto, Y., Nohara, M., Aruga-Katori, H., et al. (2007). Spin-Liquid State
in the S = 1/2 Hyperkagome Antiferromagnet Na$_{4}$Ir$_{3}$O$_{8}$. Phys. Rev.
Lett. \textbf{99}: 137207. DOI: 10.1103/PhysRevLett.99.137207.

6. Li, Y.-D., Wang, X., and Chen, G. (2016). Anisotropic spin model of strong
spin-orbit-coupled triangular antiferromagnets. Phys. Rev.  B \textbf{94}:
035107. DOI: 10.1103/PhysRevB.94.035107.

7. Lin, G., and Ma, J. (2023). Is there a pure quantum spin liquid? The
Innovation \textbf{4}: 100484. DOI: 10.1016/j.xinn.2023.100484.

8. Kitaev, A. (2006). Anyons in an exactly solved model and beyond. Ann. Phys.
\textbf{321}: 2-111. DOI: 10.1016/j.aop.2005.10.005.

9. Yokoi, T., Ma, S., Kasahara, S., et al. (2021). Half-integer quantized
anomalous thermal Hall effect in the Kitaev material candidate a-RuCl$_{3}$.
Science \textbf{373}: 568-572. DOI: 10.1126/science.aay5551.

10. Tanaka, O., Mizukami, Y., Harasawa, R., et al. (2022). Thermodynamic
evidence for a field-angle-dependent Majorana gap in a Kitaev spin liquid. Nat.
Phys. \textbf{18}: 429-435. DOI: 10.1038/s41567-021-01488-6.

11. Sears, J.A., Chern, L.E., Kim, S., et al. (2020). Ferromagnetic Kitaev
interaction and the origin of large magnetic anisotropy in $\alpha{}$-RuCl$_{3}$.
Nat. Phys. \textbf{16}: 837-840. DOI: 10.1038/s41567-020-0874-0.

12. Kasahara, Y., Ohnishi, T., Mizukami, Y., et al. (2018). Majorana
quantization and half-integer thermal quantum Hall effect in a Kitaev spin
liquid. Nature \textbf{559}: 227-231. DOI: 10.1038/s41586-018-0274-0.

13.\hspace{15pt}Jan\v{s}a, N., Zorko, A., Gomil\v{s}ek, M., et al. (2018).
Observation of two types of fractional excitation in the Kitaev honeycomb magnet.
Nat. Phys. \textbf{14}: 786-790. DOI: 10.1038/s41567-018-0129-5.

14.\hspace{15pt}Do, S.-H., Park, S.-Y., Yoshitake, J., et al. (2017). Majorana
fermions in the Kitaev quantum spin system $\alpha{}$-RuCl$_{3}$. Nat. Phys.
\textbf{13}: 1079-1084. DOI: 10.1038/nphys4264.

15. Banerjee, A., Yan, J., Knolle, J., et al. (2017). Neutron scattering in the
proximate quantum spin liquid a-RuCl$_{3}$. Science \textbf{356}: 1055-1059. DOI:
10.1126/science.aah6015.

16.\hspace{15pt}Maksimov, P.A., and Chernyshev, A.L. (2020). Rethinking
$\alpha{}$-RuC$_{l3}$. Phys. Rev.  Res. \textbf{2}: 033011. DOI:
10.1103/PhysRevResearch.2.033011.

17.\hspace{15pt}Laurell, P., and Okamoto, S. (2020). Dynamical and thermal
magnetic properties of the Kitaev spin liquid candidate $\alpha{}$-RuCl$_{3}$.
npj Quantum Mater. \textbf{5}: 1-10. DOI: 10.1038/s41535-019-0203-y.

18.\hspace{15pt}Wang, J., Normand, B., and Liu, Z.-X. (2019). One Proximate
Kitaev Spin Liquid in the K-J-$\Gamma{}$ Model on the Honeycomb Lattice. Phys.
Rev.  Lett. \textbf{123}: 197201. DOI: 10.1103/PhysRevLett.123.197201.

19.\hspace{15pt}Chaloupka, J., Jackeli, G., and Khaliullin, G. (2010).
Kitaev-Heisenberg Model on a Honeycomb Lattice: Possible Exotic Phases in Iridium
Oxides A$_{2}$IrO$_{3}$. Phys. Rev.  Lett. \textbf{105}: 027204. DOI:
10.1103/PhysRevLett.105.027204.

20.\hspace{15pt}Banerjee, A., Bridges, C.A., Yan, J.Q., et al. (2016). Proximate
Kitaev quantum spin liquid behaviour in a honeycomb magnet. Nat. Mater.
\textbf{15}: 733-740. DOI: 10.1038/nmat4604.

21.\hspace{15pt}Chaloupka, J., Jackeli, G., and Khaliullin, G. (2013). Zigzag
Magnetic Order in the Iridium Oxide Na$_{2}$IrO$_{3}$. Phys. Rev.  Lett.
\textbf{110}: 097204. DOI: 10.1103/PhysRevLett.110.097204.

22.\hspace{15pt}Jackeli, G., and Khaliullin, G. (2009). Mott Insulators in the
Strong Spin-Orbit Coupling Limit: From Heisenberg to a Quantum Compass and Kitaev
Models. Phys. Rev.  Lett. \textbf{102}: 017205. DOI:
10.1103/PhysRevLett.102.017205.

23.\hspace{15pt}Takagi, H., Takayama, T., Jackeli, G., et al. (2019). Concept
and realization of Kitaev quantum spin liquids. Nat. Rev. Phys. \textbf{1}:
264-280. DOI: 10.1038/s42254-019-0038-2.

24.\hspace{15pt}Liu, H., Chaloupka, J., and Khaliullin, G. (2020). Kitaev Spin
Liquid in 3d Transition Metal Compounds. Phys. Rev.  Lett. \textbf{125}: 047201.
DOI: 10.1103/PhysRevLett.125.047201.

25.\hspace{15pt}Hermanns, M., Kimchi, I., and Knolle, J. (2018). Physics of the
Kitaev Model: Fractionalization, Dynamic Correlations, and Material Connections.
Ann. Rev. Conden. Matter Phys. \textbf{9}: 17-33. DOI:
10.1146/annurev-conmatphys-033117-053934.

26.\hspace{15pt}Kitagawa, K., Takayama, T., Matsumoto, Y., et al. (2018). A
spin-orbital-entangled quantum liquid on a honeycomb lattice. Nature
\textbf{554}: 341-345. DOI: 10.1038/nature25482.

27.\hspace{15pt}Hwan Chun, S., Kim, J.-W., Kim, J., et al. (2015). Direct
evidence for dominant bond-directional interactions in a honeycomb lattice
iridate Na$_{2}$IrO$_{3}$. Nat. Phys. \textbf{11}: 462-466. DOI:
10.1038/nphys3322.

28.\hspace{15pt}Zhong, R., Guo, S., Xu, G., et al. (2019). Strong quantum
fluctuations in a quantum spin liquid candidate with a Co-based triangular
lattice. PNAS \textbf{116}: 14505-14510. DOI: 10.1073/pnas.1906483116.

29.\hspace{15pt}Winter, S.M. (2022). Magnetic couplings in edge-sharing
high-spin d$^{7}$ compounds. J. Phys. Mater. \textbf{5}: 045003. DOI:
10.1088/2515-7639/ac94f8.

30.\hspace{15pt}Bruin, J.A.N., Claus, R.R., Matsumoto, Y., et al. (2022).
Robustness of the thermal Hall effect close to half-quantization in
$\alpha{}$-RuCl$_{3}$. Nat. Phys. \textbf{18}: 401-405. DOI:
10.1038/s41567-021-01501-y.

31.\hspace{15pt}Czajka, P., Gao, T., Hirschberger, M., et al. (2021).
Oscillations of the thermal conductivity in the spin-liquid state of
$\alpha{}$-RuCl$_{3}$. Nat. Phys. \textbf{17}: 915-919. DOI:
10.1038/s41567-021-01243-x.

32.\hspace{15pt}Chen, L., Gu, Y., Wang, Y., et al. (2023). Large negative
magnetoresistance beyond chiral anomaly in topological insulator candidate
CeCuAs$_{2}$ with spin-glass-like behavior. The Innovation Mater. \textbf{1}:
100011. DOI: 10.59717/j.xinn-mater.2023.100011.

33. Ma, J. (2023). Spins don't align here. Nat. Phys. \textbf{19}: 922. DOI:
10.1038/s41567-023-

02041-3

34.\hspace{15pt}Yao, W., Iida, K., Kamazawa, K., and Li, Y. (2022). Excitations
in the Ordered and Paramagnetic States of Honeycomb Magnet
Na$_{2}$Co$_{2}$TeO$_{6}$. Phys. Rev.  Lett. \textbf{129}: 147202. DOI:
10.1103/PhysRevLett.129.147202.

35.\hspace{15pt}Kim, C., Jeong, J., Lin, G., et al. (2021). Antiferromagnetic
Kitaev interaction in J$_{eff }$= 1/2 cobalt honeycomb materials
Na$_{3}$Co$_{2}$SbO$_{6}$ and Na$_{2}$Co$_{2}$TeO$_{6}$. J. Phys.- Condens. Mat.
\textbf{34}: 045802. DOI: 10.1088/1361-648X/ac2644.

36.\hspace{15pt}Rau, J.G., Lee, E.K.-H., and Kee, H.-Y. (2014). Generic Spin
Model for the Honeycomb Iridates beyond the Kitaev Limit. Phys. Rev.  Lett.
\textbf{112}: 077204. DOI: 10.1103/PhysRevLett.112.077204.

37.\hspace{15pt}Winter, S.M., Li, Y., Jeschke, H.O., and Valent\'{\i}, R.
(2016). Challenges in design of Kitaev materials: Magnetic interactions from
competing energy scales. Phys. Rev.  B \textbf{93}: 214431. DOI:
10.1103/PhysRevB.93.214431.

38.\hspace{15pt}Liu, H. (2021). Towards Kitaev spin liquid in 3d transition
metal compounds. Int. J. Mod. Phys. B \textbf{35}: 21300061. DOI:
10.1142/s0217979221300061.

39.\hspace{15pt}Hong, X., Gillig, M., Hentrich, R., et al. (2021). Strongly
scattered phonon heat transport of the candidate Kitaev material
Na$_{2}$Co$_{2}$TeO$_{6}$. Phys. Rev.  B \textbf{104}: 144426. DOI:
10.1103/PhysRevB.104.144426.

40.\hspace{15pt}Chen, W., Li, X., Hu, Z., et al. (2021). Spin-orbit phase
behavior of Na$_{2}$Co$_{2}$TeO$_{6}$ at low temperatures. Phys. Rev.  B
\textbf{103}: 180404. DOI: 10.1103/PhysRevB.103.L180404.

41.\hspace{15pt}Songvilay, M., Robert, J., Petit, S., et al. (2020). Kitaev
interactions in the Co honeycomb antiferromagnets Na$_{3}$Co$_{2}$SbO$_{6}$ and
Na$_{2}$Co$_{2}$TeO$_{6}$. Phys. Rev.  B \textbf{102}: 224429. DOI:
10.1103/PhysRevB.102.224429.

42.\hspace{15pt}Bera, A.K., Yusuf, S.M., Kumar, A., and Ritter, C. (2017).
Zigzag antiferromagnetic ground state with anisotropic correlation lengths in the
quasi-two-dimensional honeycomb lattice compound Na$_{2}$Co$_{2}$TeO$_{6}$. Phys.
Rev.  B \textbf{95}: 094424. DOI: 10.1103/PhysRevB.95.094424.

43.\hspace{15pt}Lefran\c{c}ois, E., Songvilay, M., Robert, J., et al. (2016).
Magnetic properties of the honeycomb oxide Na$_{2}$Co$_{2}$TeO$_{6}$. Phys. Rev.
B \textbf{94}: 214416. DOI: 10.1103/PhysRevB.94.214416.

44.\hspace{15pt}Pilch, P., Peedu, L., Bera, A.K., et al. (2023). Field- and
polarization-dependent quantum spin dynamics in the honeycomb magnet
Na$_{2}$Co$_{2}$TeO$_{6}$: Magnetic excitations and continuum. Phys. Rev.  B
\textbf{108}: 140406. DOI: 10.1103/PhysRevB.108.L140406.

45.\hspace{15pt}Samarakoon, A.M., Chen, Q., Zhou, H., and Garlea, V.O. (2021).
Static and dynamic magnetic properties of honeycomb lattice antiferromagnets
Na$_{2}$M$_{2}$TeO$_{6}$, M = Co and Ni. Phys. Rev.  B \textbf{104}: 184415. DOI:
10.1103/PhysRevB.104.184415.

46.\hspace{15pt}Sanders, A.L., Mole, R.A., Liu, J., et al. (2022). Dominant
Kitaev interactions in the honeycomb materials Na$_{3}$Co$_{2}$SbO$_{6}$ and
Na$_{2}$Co$_{2}$TeO$_{6}$. Phys. Rev.  B \textbf{106}: 014413. DOI:
10.1103/PhysRevB.106.014413.

47. Stone, M.B., Niedziela, J.L., Abernathy, D.L., et al. (2014). A comparison
of four direct geometry time-of-flight spectrometers at the Spallation Neutron
Source. Rev. Sci. Instrum. \textbf{85}: 045113. DOI: 10.1063/1.4870050.

48.\hspace{15pt}Granroth, G.E., Kolesnikov, A.I., Sherline, T.E., et al. (2010).
SEQUOIA: A Newly Operating Chopper Spectrometer at the SNS. J. Phys.: Conf. Ser.
\textbf{251: }012058. DOI: 10.1088/1742-6596/251/1/012058.

49.\hspace{15pt}Isono, T., Kamo, H., Ueda, A., et al. (2014). Gapless Quantum
Spin Liquid in an Organic Spin-1/2 Triangular-Lattice
$\kappa{}$-H$_{3}$(Cat-EDT-TTF)$_{2}$. Phys. Rev.  Lett. \textbf{112}: 177201.
DOI: 10.1103/PhysRevLett.112.177201.

50.\hspace{15pt}Okazaki, R., Shibauchi, T., Shi, J., et al. (2011). Rotational
Symmetry Breaking in the Hidden-Order Phase of URu$_{2}$Si$_{2}$. Science \textbf{331}: 439-442.
DOI: 10.1126/science.1197358.

51.\hspace{15pt}Leahy, I.A., Pocs, C.A., Siegfried, P.E., et al. (2017).
Anomalous Thermal Conductivity and Magnetic Torque Response in the Honeycomb
Magnet $\alpha{}$-RuCl$_{3}$. Phys. Rev.  Lett. \textbf{118}: 187203. DOI:
10.1103/PhysRevLett.118.187203.

52.\hspace{15pt}Asaba, T., Lawson, B.J., Tinsman, C., et al. (2017). Rotational
Symmetry Breaking in a Trigonal Superconductor Nb-doped Bi$_{2}$Se$_{3}$. Phys.
Rev.  X \textbf{7}: 011009. DOI: 10.1103/PhysRevX.7.011009.

53.\hspace{15pt}Lee, C.H., Lee, S., Choi, Y.S., et al. (2021). Multistage
development of anisotropic magnetic correlations in the Co-based honeycomb
lattice Na$_{2}$Co$_{2}$TeO$_{6}$. Phys. Rev.  B \textbf{103}: 214447. DOI:
10.1103/PhysRevB.103.214447.

\vspace{8pt}
\textbf{ACKNOWLEDGEMENTS}
\vspace{8pt}

We gratefully acknowledge the helpful discussions of Dr. Tian Shang, East China
Normal University. J.M. and Z.X.L. thank the financial support from the National
Key Research and Development Program of China (Grant Nos. 2022YFA1402702,
2018YFA0704300, and 2023YFA1406500). G.T.L, Z.X.L., and J.M. thank the National
Science Foundation of China (Nos. U2032213, 11774223, 12004243, 11974421,
12374166 and 12134020). J.M. thanks the interdisciplinary program Wuhan National
High Magnetic Field Center (Grant No. WHMFC 202122), Huazhong University of
Science and Technology, and a Shanghai talent program. G.T.L thanks the projects
funded by China Postdoctoral Science Foundation (Grant No. 2022T150414) and the
Startup Fund for Young Faculty at SJTU (24X010500168). Q.H. and H.D.Z. thank the
support from NSFDMR-2003117. M.F.S thanks the support from Guangdong Provincial
Key Laboratory of Extreme Conditions (Grant No. 2023B1212010002). H.W.C. thanks
the support from the Collaborative Innovation Program of Hefei Science Center,
CAS (Grants No. 2021HSCKPRD003). This research used resources at the Spallation
Neutron Source, a DOE Office of Science User Facility operated by the Oak Ridge
National Laboratory. The work at Michigan is supported by the Department of
Energy under Award No. DE-SC0020184 (magnetic torque analysis).

\vspace{8pt}
\textbf{AUTHOR CONTRIBUTIONS }
\vspace{8pt}

G.T. Lin, and J. Ma conceived and supervised the study and designed the
measurement setup. Q. Huang and H.D. Zhou synthesized the high-quality
single-crystal samples. G.T. Lin, M.F. Shu, Y.N.N. Ma, J.L. Jiao, J.M. Sheng,
L.S. Wu, L. Li and G. Li performed the magnetization, heat capacity, and torque
measurements. A. Kolesnikov, G.T. Lin, and J. Ma performed inelastic neutron
scattering experiment. G.T. Lin, X.Q. Wang, J. Ma, H.D. Zhou, and Z.X. Liu
analyzed the data. Theoretical interpretations and numerical simulations are
carried out by Q.R. Zhao, G.J. Duan, R. Yu and Z.X. Liu. G.T. Lin prepared the
manuscript with H.D. Zhou, Z.X. Liu and J. Ma. All authors discussed the data and
its interpretation.

\pagestyle{empty}
\setcounter{equation}{0}
\renewcommand{\theequation}{S\arabic{equation}}
\setcounter{figure}{0}
\renewcommand{\thefigure}{S\arabic{figure}}
\setcounter{section}{0}
\renewcommand{\thesection}{S\arabic{section}}
\setcounter{table}{0}
\renewcommand{\thetable}{S\arabic{table}}

\onecolumngrid


\newpage
\begin{center}
{\LARGE \textbf{Supplemental Information}}
\end{center}

\section{Honeycomb lattice}

\begin{figure*}[htbp]
\includegraphics[width=.8\linewidth]{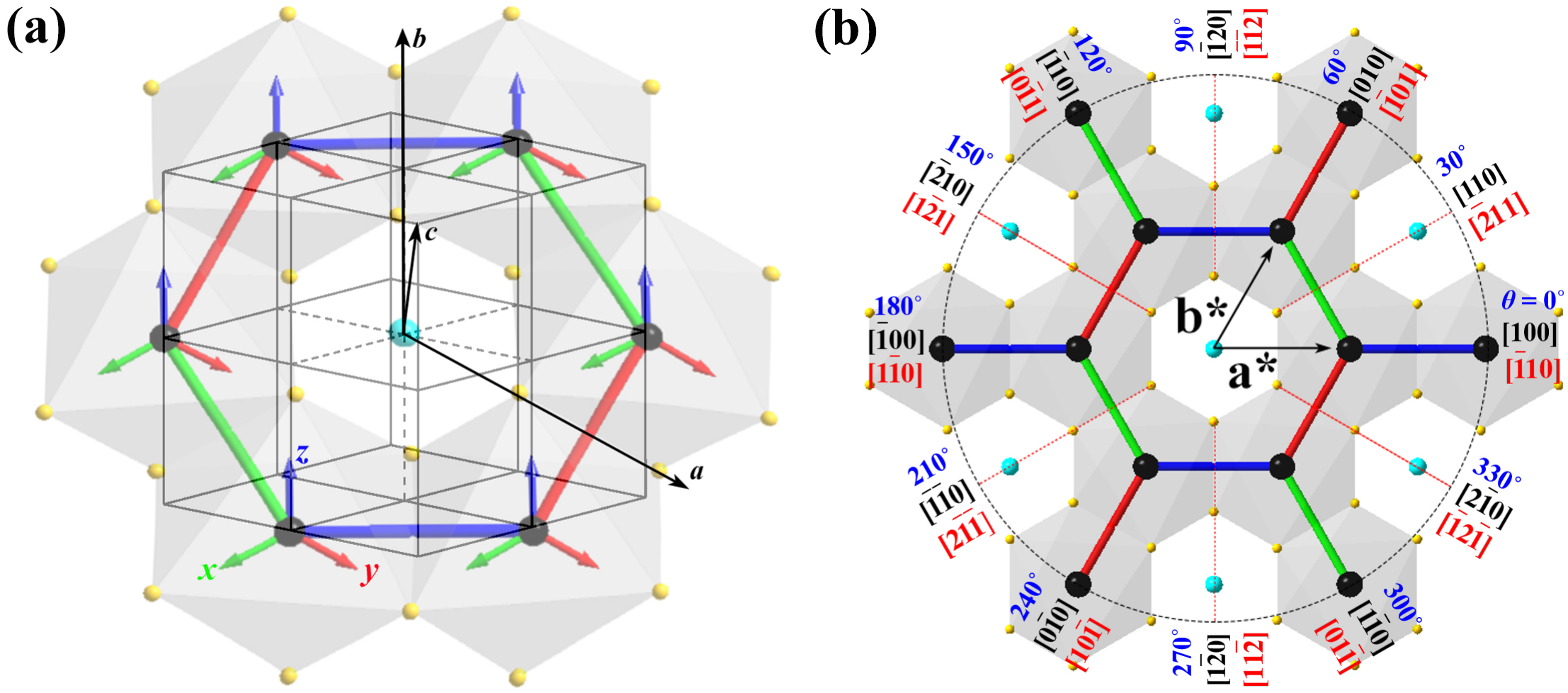}
\caption{$\textbf{Schematic of the crystal structure of Na$_2$Co$_2$TeO$_6$.}$ (A) Definitions of the crystallographic axes ($\textit{a}$, $\textit{b}$, $\textit{c}$) and the Ising spin axes ($\textit{x}$, $\textit{y}$, $\textit{z}$). (B) Honeycomb network showing the Co-Co bonds (red/blue/green) and CoO$_6$ octahedras (the black, cyan and golden spheres represent Co, Te and O atoms, respectively). The black arrows are the two principal reciprocal vectors \textbf{a$^{*}$} and \textbf{b$^{*}$} in first Brillouin zone. Magnetic field $\mathbf{B}$ is applied within the honeycomb plane. $\theta$ is the angle between $\mathbf{B}$ and $\textbf{a$^{*}$}$ and with respect to the real-space orientation of the Co-Co bonds. The black and red fonts represent the coordinate vector in reciprocal space and spin-axes coordinate, respectively. Directions that are equivalent to [1,0,0] and [1,1,0] correspond to angles $\theta$ = $\textit{n}$ $\times$ 60$^{\circ}$ and $\theta$ = $\textit{n}$ $\times$ 60$^{\circ}$ + 30$^{\circ}$, respectively (where \textit{n} is an integer) in reciprocal space coordinate.}\label{honeycomb}
\end{figure*}

The geometry of the Honeycomb layer in NCTO is shown in Fig.\ref{honeycomb}, where  Fig.\ref{honeycomb}A illustrates the bases $\bf a, b$ of translation as well as the spin axes $x,y,z$, and  Fig.\ref{honeycomb}B shows the coordinates of various directions in the ${\bf a^*, b^*, c^*}={2\pi\over{|{\bf c}|^2}}{\bf c}$ bases (black) and in the $x,y,z$ bases (red) respectively. 

\section{Magnetization and specific heat}

\begin{figure*}[htbp]
\includegraphics[width=.8\linewidth]{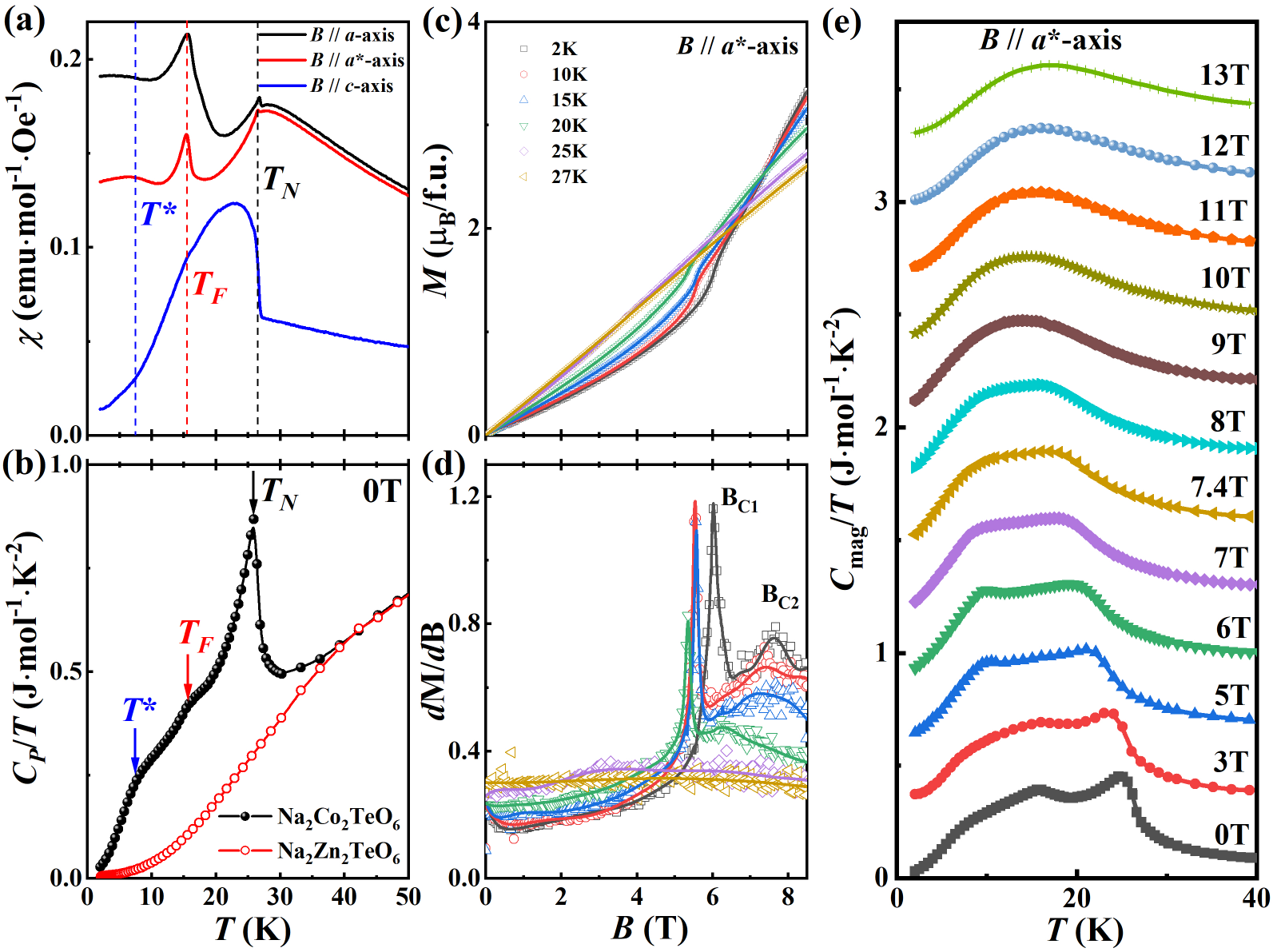}
\caption{$\textbf{Macroscopic physical properties of Na$_2$Co$_2$TeO$_6$.}$ (A) Temperature dependence of susceptibility $\chi$($\textit{T}$) in NCTO measured at 0.01 T along the different crystal-axis directions. (B) Temperature dependence of specific heat under 0 T for NCTO and Na$_2$Zn$_2$TeO$_6$. (C) The isothermal magnetization $\textit{M}$($\textit{B}$) with the applied magnetic field $\mathbf{B}$ // $\textbf{a$^{*}$}$-axis at selected temperatures. (DD) The differential isothermal magnetization as functions of fields $\textit{dM}$/$\textit{dB}$ vs. $\textit{B}$, originating from the data of (C). The solid lines represent the fitted curves by polynomial in (C) and (D). The critical fields $\textit{B}_{C1}$ and $\textit{B}_{C2}$ are determined by the $\textit{dM}$/$\textit{dB}$ curves, which are comparable with the critical fields in the main text. (E) Temperature dependence of magnetic specific heat in NCTO under selected fields.}\label{mag}
\end{figure*}

The single-crystal DC magnetic susceptibility $\chi$($\textit{T}$), Fig.~\ref{mag}A, measured by the zero-field cooling at 0.01 T and the single-crystal specific heat $\textit{C}$$_P$($\textit{T}$)/$\textit{T}$ at 0 T, Fig.~\ref{mag}B, show two sharp anomalies at $\textit{T}$$_N \approx$ 26 K with zigzag antiferromagnetic order and $\textit{T}$$_F \approx$ 15 K with a spin reorientation transition in the $\textit{ab}$ plane. A weak anomaly of the $\chi$($\textit{T}$) and $\textit{C}_P$($\textit{T}$)/$\textit{T}$ curves can be also seen around $\textit{T}^{*}  \approx$ 7 K, which may be related to the modulation of magnetic domain. Their values mirror well the previous reports.\upcite{1,2,3,4,5} Due to the magnetic-exchange anisotropy in NCTO, the magnetic order state exhibits a distinct magnetic anisotropic behavior along the different crystal-axis directions, Fig.~\ref{mag}A. However, the $\chi$($\textit{T}$) curves also present a strongly anisotropic behavior above $\textit{T}_N$ in the $\textit{ab}$ plane and $\textit{c}$-axis, which is mainly attributed to the strongly anisotropic $\textit{g}$ factor originating from the low-spin state of Co$^{2+}$. Hence, the spin-1/2 model is an effective, low-energy description for the Co$^{2+}$ ions in the temeprature range considered in this work, which is consistent with our previous report.\upcite{5}

\begin{figure*}[htbp]
\includegraphics[width=.55\linewidth]{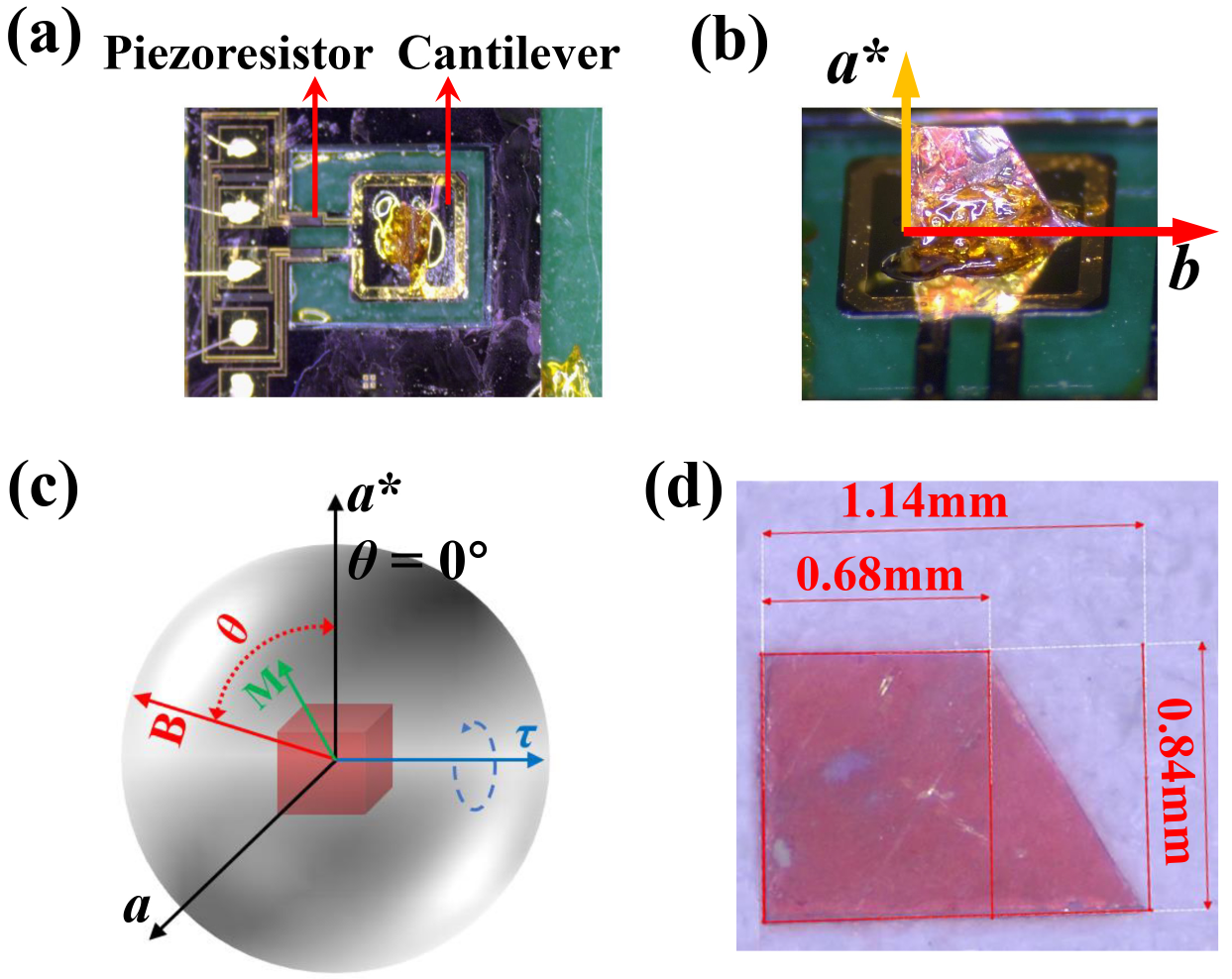}
\caption{$\textbf{Set up of magnetic torque measurements.}$ (A) and (B) illustrate the experimental configuration for angle-dependent magnetic torque $\tau$($\theta$) measurements. (C) The polar angular relations between the $\mathbf{a^{*}}$ and $\mathbf{B}$ in the rotated plane. (D) Morphology and size of the selected sample.}\label{sample}
\end{figure*}

In order to better clarify the magnetic phase diagram in the main text, we perform isothermal magnetization \textit{M}(\textit{B}) with the applied magnetic field $\mathbf{B}$ // $\textbf{a$^{*}$}$-axis at selected temperatures, Fig.~\ref{mag}C-D. The critical fields \textit{B}$_{C1}$ and \textit{B}$_{C2}$ are comparable with results of the magnetic torque.

Figure~\ref{mag}(E) shows the magnetic specific heat \textit{C}$_{mag}$/$\textit{T}$ curves with $\mathbf{B}$ // $\textbf{a$^{*}$}$-axis, subtracting phonon contributions measured on an isostructural Na$_2$Zn$_2$TeO$_6$ reference crystal that is shown in the Fig.~\ref{mag}B. The critical temperatures in Fig. 1(A) in the main text is marked by the anomalies in the \textit{C}$_{mag}$/$\textit{T}$ curves.

\begin{figure*}[t]
\includegraphics[width=.9\linewidth]{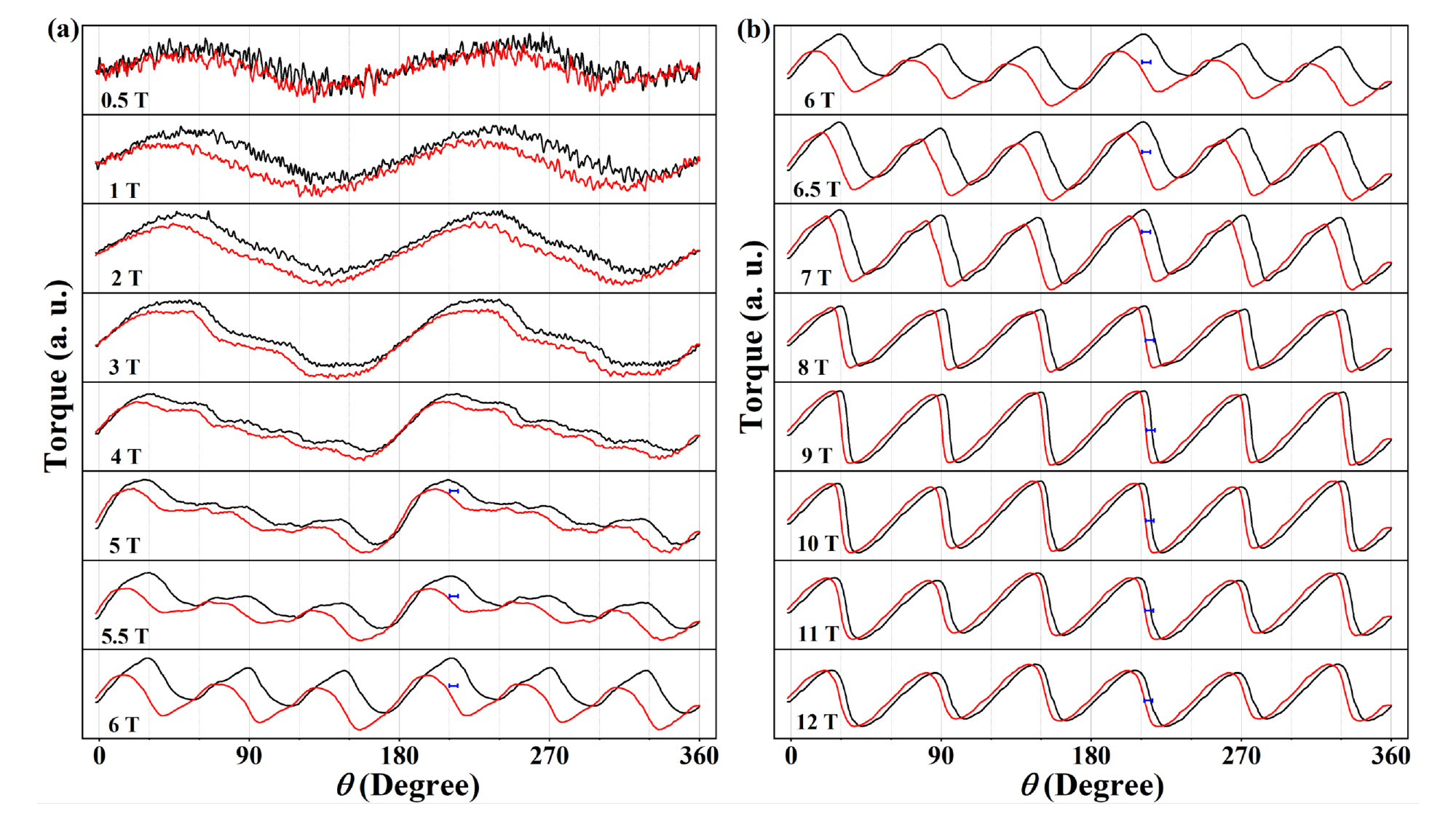} 
\caption{$\textbf{Angle-dependent magnetic torque without subtracting background at 2 K with different fields.}$ The blue horizontal error bars indicate the instrumental backlash ($\sim$ 5 degrees) when reversing rotation. The black solid lines represent the increasing angles from 0 to 360 degrees. The red solid lines represent the decreasing angles from 360 to 0 degrees.}\label{sample torque}
\end{figure*}

\begin{figure*}[htbp]
\includegraphics[width=.7\linewidth]{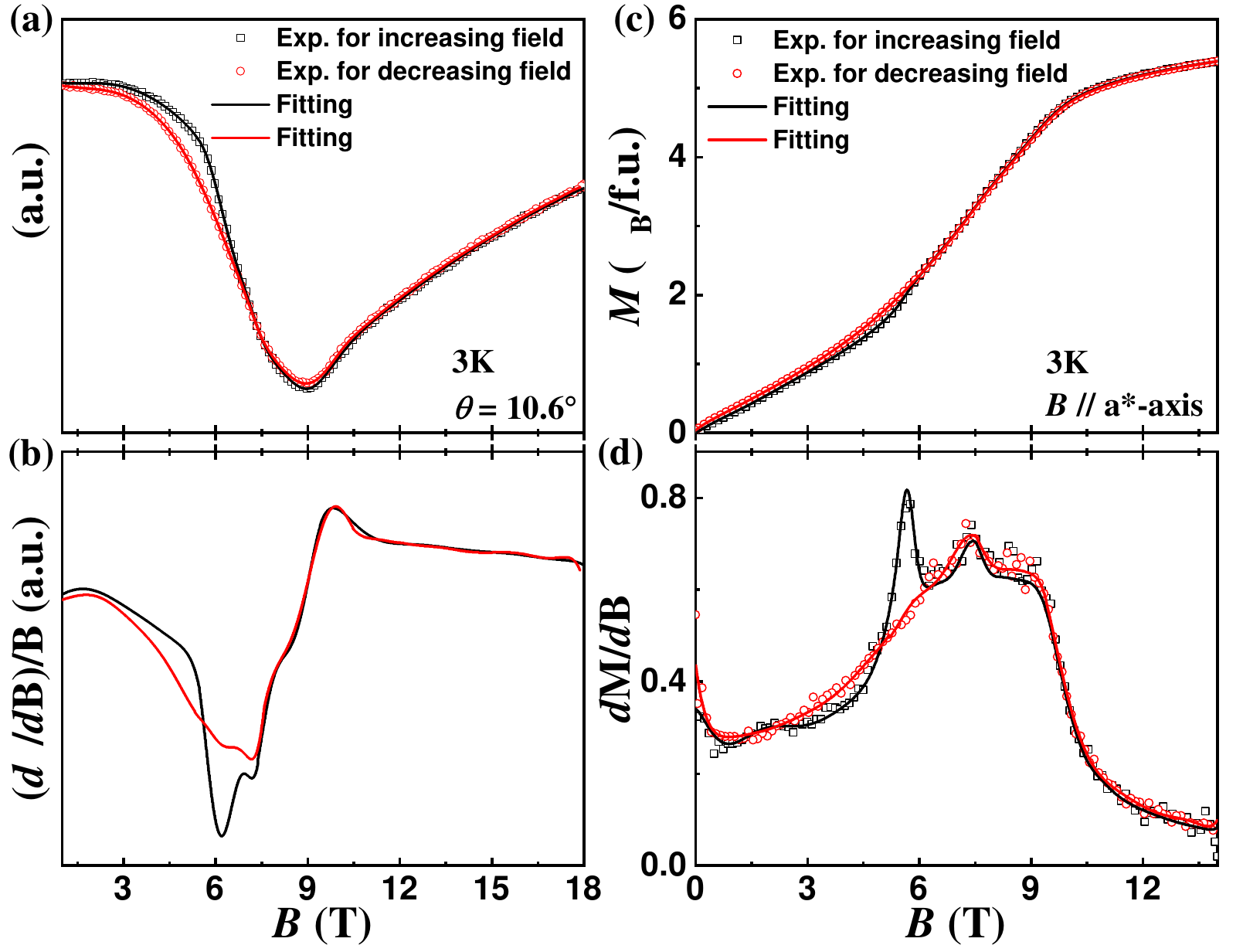}
\caption{$\textbf{Field dependence of magnetic torque and magnetization.}$ Field dependence of magnetic torque (A) and differential magnetic torque (B) in NCTO measured at 3 K with magnetic field 10.6 degrees away from $\mathbf{a}^{*}$-aixs. Field dependence of magnetization (C) and differential magnetization (D) in NCTO measured at 2 K along the field $\mathbf{B}$ // $\mathbf{a}^{*}$-axis.}\label{first}
\end{figure*}

\begin{figure*}[htbp]
\includegraphics[width=.9\linewidth]{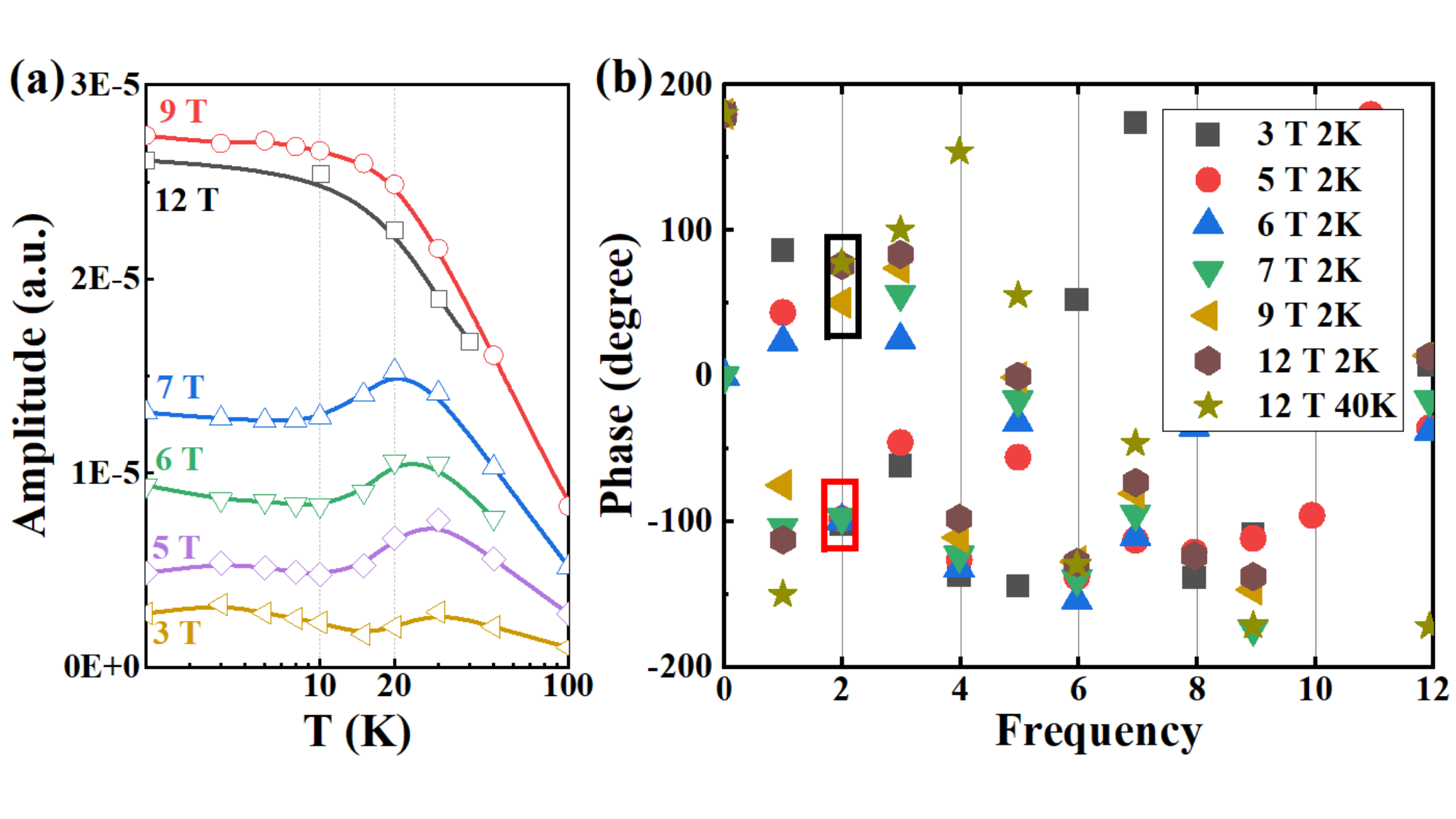}
\caption{$\textbf{Fourier transform of angle-dependent magnetic torque.}$ (A) Temperature-dependent amplitude of 2-fold symmetry obtained by Fourier transform of angle-dependent magnetic torque at different fields. (B) The frequency dependence of phases by the Fourier Transform from the angle-dependent torque. The 2-fold symmetric structure from the magnetic order labeled by the red box. The 2-fold symmetric structure from the sample misalignment marked by the black box.}\label{FFT of C2}
\end{figure*}

\begin{figure*}[htbp]
\includegraphics[width=.7\linewidth]{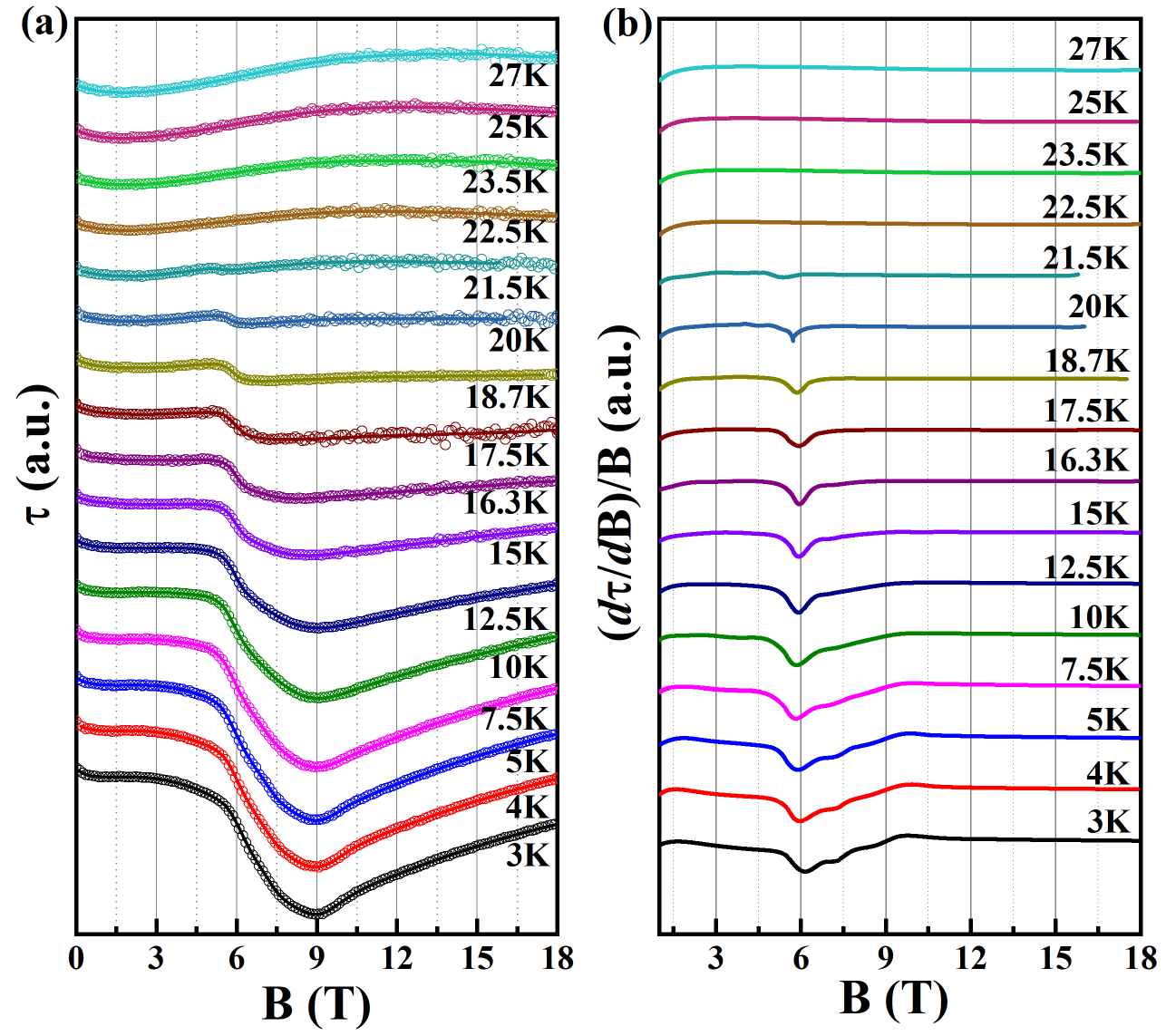}
\caption{$\textbf{Isothermal magnetic torque with field close to $\mathbf{a}^{*}$-axis.}$  Field dependence of magnetic torque (A) and differential magnetic torque (B) in NCTO measured at different temperatures along the angle $\theta$ = 10.6°.}\label{sample tq}
\end{figure*}

\begin{figure*}[htbp]
\includegraphics[width=.6\linewidth]{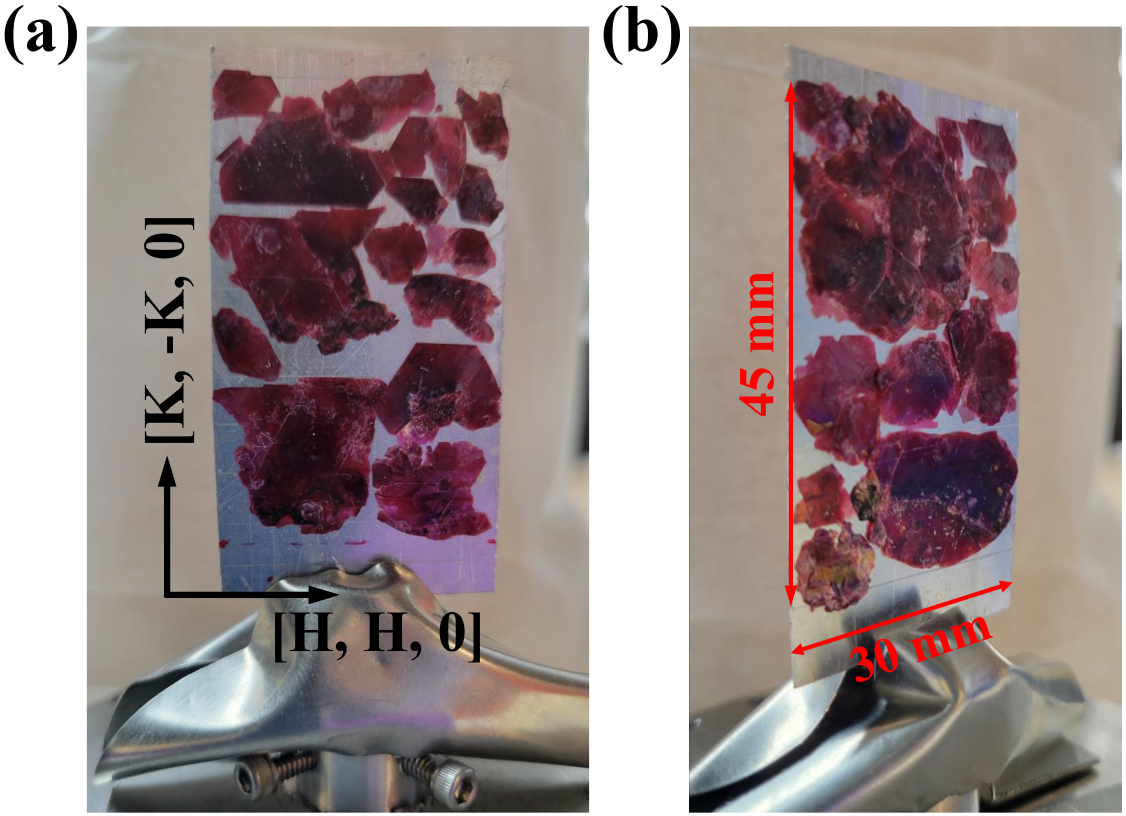}
\caption{$\textbf{The picture of the co-aligned NCTO single crystals for INS experiment}$ (A) and (B) Single-crystal samples used in the SEQUOIA time-of-flight spectrometer at the spallation neutron source, Oak Ridge National Laboratory, USA. The co-aligned samples were adhered to both sides of the aluminum sheet shown in the (A) and (B).}\label{INS sample}
\end{figure*}

\section{Magnetic torque}

Figure~\ref{sample torque} shows the angle-dependent magnetic torque at 2 K. The backlash of the mechanical rotation is about 5 degree, marked by the blue horizontal error bars in Fig.~\ref{sample torque}. Therefore, we observe that the angle-dependent magnetic torque curves will have hysteresis obtained by counterclockwise and clockwise rotations. Surprisingly, a larger direction deviation of more than 5 degrees is seen between about 5 T and 7 T, which cannot be attributed to the error caused by the instrument. We note that an obvious hysteresis can be seen in the field dependence of magnetic torque and magnetization, Fig.~\ref{first}, indicating the first-order phase transition around \textit{B}$_{C1}$ that is just between 5 T and 7 T. Hence, such large angular deviation of Fig.~\ref{sample torque} between about 5 T and 7 T should be related to the first-order phase transition around \textit{B}$_{C1}$.

Actually, the ‘X phase’ is not “another disordered phase”. Instead, according to our experimental data and theoretical calculation, we identify it with a distinct phase with coexisting zigzag order phase and {\R Z$_2$} topological order. The {\R Z$_2$} topological order is reflected by the fact that there is a four-fold ground state degeneracy on a torus, indicating the deconfined {\R Z$_2$} gauge fluctuations.

Firstly, the transition from the ordered phase to the intermediate disordered phase is of first order (which is consistent with our theory). The evidence for this first order phase transition is provided in the supplemental material, where obvious hysteresis structure is observed in the torque data. In Fig.~\ref{sample torque}, the angular dependence of the torque curves shows a big hysteresis with field around 6 T when the sample (or equivalently the field direction) is rotating clockwise or counterclockwise. Furthermore, in  Fig.~\ref{first}, the field-dependent torque curves (with fixed field intensity) also have a hysteresis structure around 6T (with fixed angle) when the field strength is increasing or decreasing.

Secondly, the angular dependence of the torque shows the same (the phases of the 2-period Fourier components are the same, see Fig.~\ref{FFT of C2}) 2-fold symmetric structure for field intensities B=3T, 6T, 7T, indicating the existence of zigzag order up to 7T. On the other hand, when the field strength exceeds 6T, the torque data eventually establish a 6-fold symmetry, indicating the appearance of disordered state. Especially, between \textit{B}$_{C1}$=6T and \textit{B}$_{C2}$=7.5T, the angular dependence of the torque contains both 2-fold and 6-fold symmetric components.

Thirdly, the first-order nature of the transition between the magnetically ordered phase and the disordered phase had also been observed in previous studies.\upcite{5,6}

From the above three facts, the most reasonable possibility is that between \textit{B}$_{C1}$ and \textit{B}$_{C2}$ is a phase with coexisting zigzag order and strong QSL fluctuations. Indeed, this phase is very interesting and to further confirm the coexistence of the two types of orders in this field region, more experimental explorations are needed. But this is not the focus of the present study. Our main interest is focused on the field-induced QSL phase between \textit{B}$_{C2}$ and \textit{B}$_{C3}$.

In order to draw the magnetic phase diagram in the Fig. 1(c) of the main text, we measure the field dependence of magnetic torque along the angle $\theta$ = 10.6$^{\circ}$ at different temperatures, Fig.~\ref{sample tq}. The superimposed color map of Fig. 1(c) is from the differential ${1\over B}{d\tau\over dB}$ of Fig.~\ref{first}.

\section{Inelastic neutron scattering}

\begin{figure*}[htbp]
\includegraphics[width=10cm]{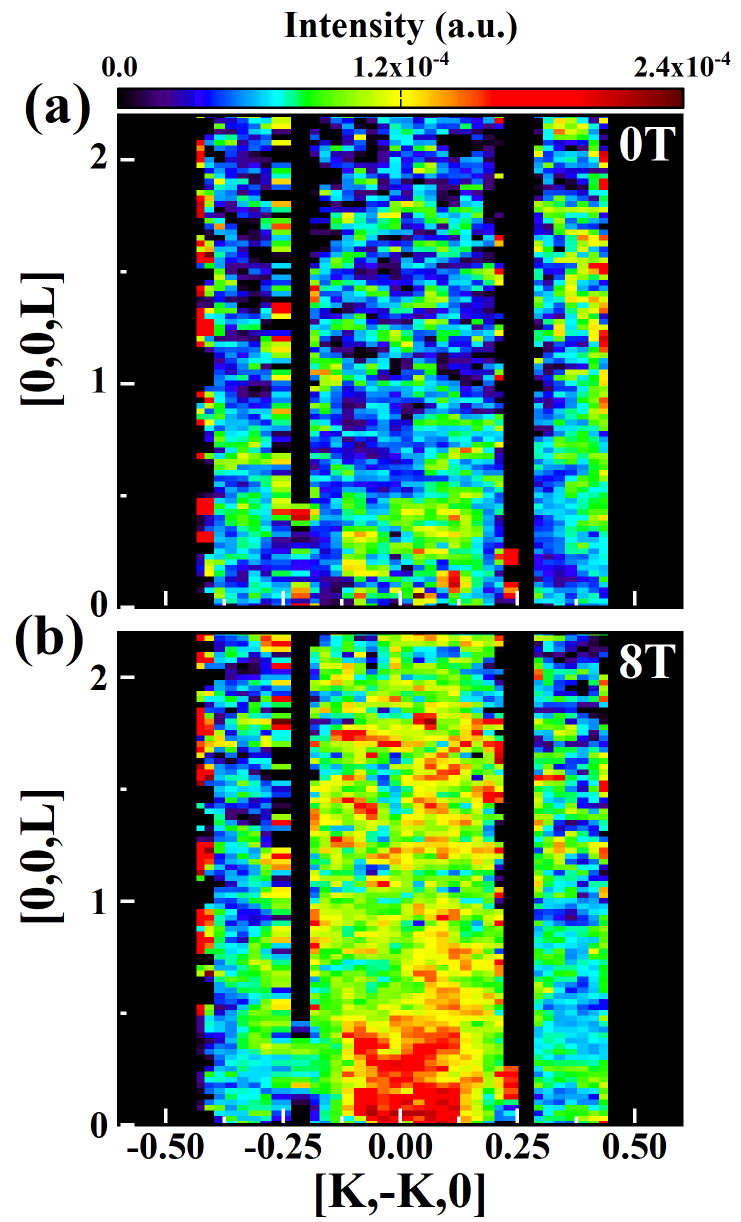}
\caption{$\textbf{Rod-like dependence on the out-of-plane momentum component L.}$ (A) and (B) Constant-energy scattering at 0 T and 8 T, respectively, integrated over H = [–0.1, 0.1] and E = [1.5, 2.5] meV, projected on the reciprocal honeycomb plane defined by the perpendicular directions [0, 0, L] and [K, –K, 0]. The color bar indicates scattering intensity in arbitrary unit in linear scale. The black regions lack detector coverage.}\label{INS}
\end{figure*}

In the main text, the data of Fig. 1D-E and Fig. 3A-D are collected from SEQUOIA time-of-flight spectrometer. In all these experiments, the single crystals were co-aligned in the (HHL) scattering plane with $\mathbf{B}$ // $\textbf{a$^{*}$}$-axis, Fig.~\ref{INS sample}A. About 0.559 g samples were fixed on an aluminum sheet with 30 × 60 × 0.5 mm$^3$ in size and the effective area of samples is about 30 × 45 mm$^2$, Fig.~\ref{INS sample}B. The software suite DAVE was used to visualize the time-of-flight neutron scattering data and the space group P6$_3$22 with a = b = 5.227 \AA, c = 11.2231 \AA was also adopted to analyze these INS results.

Figures~\ref{INS}A-B show the rod-like dependence on the out-of-plane momentum component L, demonstrating the inter-layer interaction is relatively weak, NCTO could be regarded as a 2D Co$^{2+}$-based honeycomb magnet.

\section{Variational Monte Carlo Simulations}

\subsection{For the Ground States} %

The VMC approach is based on the spinon representation by introducing two species of fermions $C_i^\dagger = (c_{i\uparrow}^\dagger,c_{i\downarrow}^\dagger)$. The spin operators are written in quadratic forms of the fermionic spinons $S_i^m =\frac{1}{2} C_i^\dagger \sigma^m C_i$ under the particle number constraint, $\hat{N_i} = c_{i\uparrow}^\dagger c_{i\uparrow} + c_{i\downarrow}^\dagger c_{i\downarrow} = 1$, where $m \equiv x,y,z$, and $\sigma^m$ is Pauli matrix. It is convenient  to introduce the matrix operator $\psi_i=( C_i, \bar C_i)$ with $\bar C_i=(c_{i\dn}^\dag, -c_{i\up}^\dag)^T$ such that the spin operators can also be written as $S_i^m = {\rm Tr}(\psi_i ^\dag {\sigma^m \over4}\psi_i)$. Since the spin operators are invariant under a local SU(2) transformation $\psi_i\to \psi_i W_i$, the fermionic spinon representation has an SU(2) gauge `symmetry'.\upcite{Anderson88}

The spin interactions are rewritten in terms of interacting fermionic operators and are further decoupled into a non-interacting mean-field Hamiltonian.  In a quantum spin liquid (QSL) phase, all the physical symmetries are preserved in the ground state, and the SU(2) gauge symmetry may reduce to U(1) or Z$_2$. For instance, the gauge symmetry of the Kitaev spin liquid is $Z_2$. The mean-field Hamiltonian of a QSL phase does not necessarily preserve the physical symmetries, but should be invariant under an extended group called the 'projective symmetry group' (PSG).\upcite{Wen02} We adopt the same PSG of the Kitaev spin liquid,\upcite{You12, Wang19} and the corresponding QSL mean field Hamiltonian of the $K$-$J_1$-$\Gamma$-$\Gamma'$-$J_3$  model reads
\beq
H_{\rm mf}^{\rm QSL} & = & \sum_{\langle i,j \rangle\in\gamma} \left[ i \rho_a {\rm Tr} (\psi_i^\dagger \psi_j + \tau^x \psi_i^\dagger \sigma^x \psi_j + \tau^y \psi_i^\dagger \sigma^y \psi_j + \tau^z \psi_i^\dagger \sigma^z \psi_j)\right. \notag\\
&&\left. + i \rho_c {\rm Tr} (\psi_i^\dagger \psi_j + \tau^\gamma \psi_i^\dagger \sigma^\gamma \psi_j - \tau^\alpha \psi_i^\dagger \sigma^\alpha \psi_j - \tau^\beta \psi_i^\dagger \sigma^\beta \psi_j) 
+i\rho_d {\rm Tr} ( \tau^\alpha \psi_i^\dag \sigma^\beta \psi_j + \tau^\beta \psi_i^\dag \sigma^\alpha \psi_j ) \right.\notag\\
&&\left. + i\rho_f {\rm Tr} (\tau^\alpha \psi_i^\dag \sigma^\gamma \psi_j + \tau^\gamma \psi_i^\dag
\sigma^\alpha \psi_j   + \tau^\beta \psi_i^\dag \sigma^\gamma \psi_j + \tau^\gamma \psi_i^\dag \sigma^\beta \psi_j )+  {\rm H.c.}  \right] \notag\\
&&+ \sum_{\langle\langle\langle i,j\rangle\rangle\rangle}\left[ t_3 {\rm Tr} (\tau^z\psi_i^\dag \psi_j) + \Delta_3  {\rm Tr} (\tau^x \psi_i^\dag \psi_j) +  {\rm H.c.} \right]  + \sum_{i} \pmb{\lambda}_{i} \cdot  \operatorname{Tr}(\psi_{i} \pmb{\tau} \psi_{i}^{\dagger})
\eeq
where $\rho_a$, $\rho_c$, $\rho_d$, and $\rho_f$ come from the nearest neighbor interactions, $t_3$ and $\Delta_3$ come from the $J_3$ Heisenberg exchange interactions, $\sigma^m$ stands for the spin operation, $\tau^m$ denotes the gauge operation, and $\pmb \lambda$ is imposed to ensure the SU(2) gauge invariance (the $\lambda^z$ component is the Lagrangian multiplier for the particle number constraint).

Furthermore, to describe the long-range magnetic order, we introduce a background field $\pmb M_i$ to enforce the symmetry-breaking. The ordering pattern  is contained by the single-$\pmb Q$ ansatz\upcite{HYKee14} with
$$\pmb{M}_i = M \Big( \sin \phi \big[\hat {\pmb e}_x \cos (\pmb{Q} \cdot
\pmb{r}_i) + \hat{\pmb e}_y \sin(\pmb{Q} \cdot \pmb{r}_i)\big] + \cos \phi
\, \hat{\pmb e}_z \Big ),$$
where $\pmb Q$ is the ordering momentum, $\hat {\pmb e}_{x,y,z}$ are the local spin axes, and $\phi$ is the canting angle. The ordering momentum $\pmb Q$ is adopted either from the classical ground state.
For a given $\pmb Q$, the local axes $\hat {\pmb e}_{x,y,z}$ are fixed as they are in the classical state, while $M$ and $\phi$ are treated as variational parameters.

Hence, the complete trial mean-field Hamiltonian for the $K$-$J_1$-$\Gamma$-$\Gamma'$-$J_3$  model in an external magnetic field reads
\begin{equation}\label{Order}
H_{\rm mf}^{\rm total} = H_{\rm mf}^{\rm QSL} + {\textstyle \frac{1}{2}} \sum_i
(\pmb {M}_i   + \tilde g\mu_B \tilde {\pmb B}) \cdot  {\rm Tr} ( \psi_i^\dag {\pmb \sigma\over2}\psi_i), 
\end{equation}
where $\tilde {\pmb B}$  is the effective Zeeman field due to the external magnetic field $\pmb B$, but $\tilde {\pmb B}$ is not necessarily equal to $\pmb B$.

 Then we perform Gutzwiller projection to the mean-field ground state $|\Psi_{\rm mf} (\pmb R)\rangle$ to enforce the particle number constraint. The projected states $|\Psi (\pmb R)\rangle = P_G |\Psi_{\rm mf}(\pmb R) \rangle$ provide a series of trial wave functions depending on the choice of the mean-field Hamiltonian $H_{\rm mf}(\pmb R)$, where $P_G$ denotes a Gutzwiller projection and $\pmb R$ are treated as variational parameters.  The energy of the trial state $E (\pmb R) = \langle \Psi(\pmb R) |H| \Psi(\pmb R) \rangle / \langle \Psi(\pmb R)| \Psi(\pmb R) \rangle$ is computed using Monte Carlo sampling, and the optimal parameters $\pmb R$ are determined by minimizing the energy $E(\pmb R)$.

%
%
%
%
%

\begin{figure}[b]
\subfigure[ phase diagram with $B\parallel(\hat x-\hat y)$]{\includegraphics[width=10.0cm]{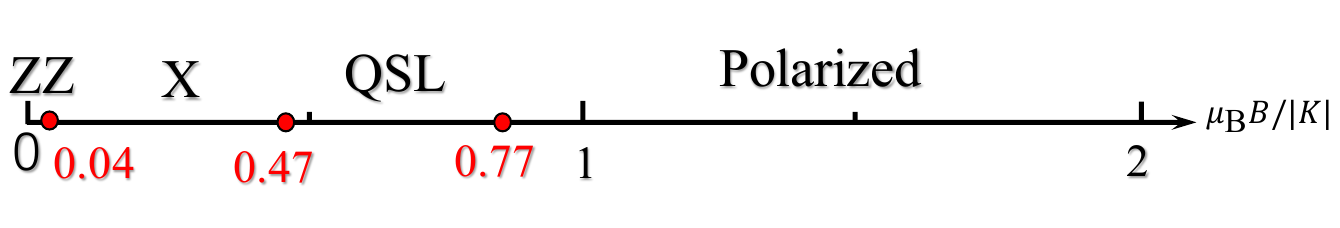}}
\subfigure[ spinon dispersion in the QSL phase]{\includegraphics[width=10.0cm]{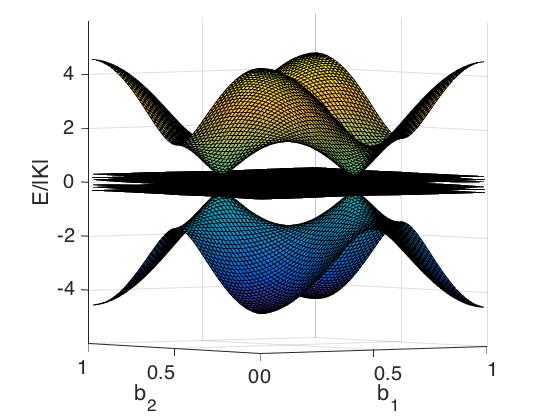}}
\caption{\textbf{(A) Calculated phase diagram with varying field strength and (B) the spinon dispersion in the QSL phase.} The field is applied along the ${\bf a}^*={1\over\sqrt2}({\hat x-\hat y})$-direction, and the in-plane Land$\Acute{e}$ $g$ factor is adopted as $g=4.13$.\upcite{5}}\label{phasediag}
\end{figure}

In the variational process, we first choose a classical metastable configuration with $\pmb Q$, then optimize the energy of the projected state. By comparing the optimal energies of different trial $\pmb Q$s, we obtain the approximate ground state. When the variational parameter $\pmb M_i$ is zero, the magnetic order disappear, and system goes into QSL phase.

\subsubsection{Magnitude dependence of the magnetic field} 
The phase diagram of the $K$-$J_1$-$\Gamma$-$\Gamma'$-$J_3$  model with $K=1.408$meV, $J_1/|K|=-1.09, \Gamma/|K|=-0.94, \Gamma'/|K|=0.63, J_3/|K|= 0.94$ under applied in-plane magnetic field applied along the ${\bf a}^*={1\over\sqrt2}(\hat x-\hat y)$-direction is shown in Fig.~\ref{phasediag}(A), where four phases appear. The first phase is a classical phase with long-ranged zigzag order, the third phase and the fourth phase are the QSL phase and the trivial polarized phase, respectively. The second phase is an interesting phase with coexisting zigzag type magnetic order and $Z_2$ topological order (the QSL phase is also topologically ordered). The $Z_2$ topological order is reflected by the four-fold ground state degeneracy on a torus.

{\it Remarks: The adopted interaction parameters in this work are proportional to the tx+ set of parameter in Ref.\upcite{Rachel21}, namely, $J_{1}=$-3.5meV, $K=$3.2meV, $\Gamma=$-3.0meV, $\Gamma'=$2meV, excepted that the $J_3$ is enlarged to 3meV  to ensure that the spin wave dispersion in the disordered phase is of concave shape at the $\Gamma$ point (see section \ref{sec:SW}). Furthermore, to compare the excitation spectrum (see section \ref{sec:DSF}) with experiment, we then divide all the parameters by a factor 2.5. Comparing the VMC result with the phase diagram of Fig.1(a) in the main text, the theoretically obtained critical magnetic fields in Fig.\ref{phasediag}, namely$B_{C2} = 0.47|K|/\mu_{B} \approx 11.4$T and $B_{C3} = 0.77|K|/\mu_{B} \approx 18.7$T are slightly larger than the experimental values.
This is possibly caused by the simplified $K$-$J_1$-$\Gamma$-$\Gamma'$-$J_3$ effective model where only two-body interactions are considered. The ignored multi-spin interactions 
may suppress the zigzag order and reduce the critical magnetic field.} 

Noticing that the partially polarized QSL phase and the polarized trivial phase have the same symmetry. The difference is that the QSL phase is $Z_2$ deconfined such that the spions and gauge fluxes are elementary excitations and the polarized trivial phase is $Z_2$ confined such that the elementary excitations are magnons. The deconfinement of the $Z_2$ gauge field can also be reflected from the degeneracy of the ground states on a torus. If the $Z_2$ gauge field is deconfined, then $Z_2$ fluxes are allowed low energy excitations. Consequently, there will be four-fold degenerate ground states on a torus (given that the Chern number of the mean field ground state is zero).  On the other hand, if the $Z_2$ gauge field is confined, then there will be a single ground state on a torus.

The four-fold degeneracy in the $Z_2$ deconfined QSL phase can be understood as the following. Notice that the interactions are short-ranged, the energy density is a local quantity, but the global fluxes in the two holes are non-local. Hence in the thermodynamic limit the inserting of $\pi$-flux in one of the holes does not change the energy density of the system. The existence or absence of $Z_2$ fluxes through the two holes results in four-fold degenerate ground states on the torus.

\begin{figure*}[b]
\subfigure[GSD in the QSL phase]{
\includegraphics[width=7.0cm]{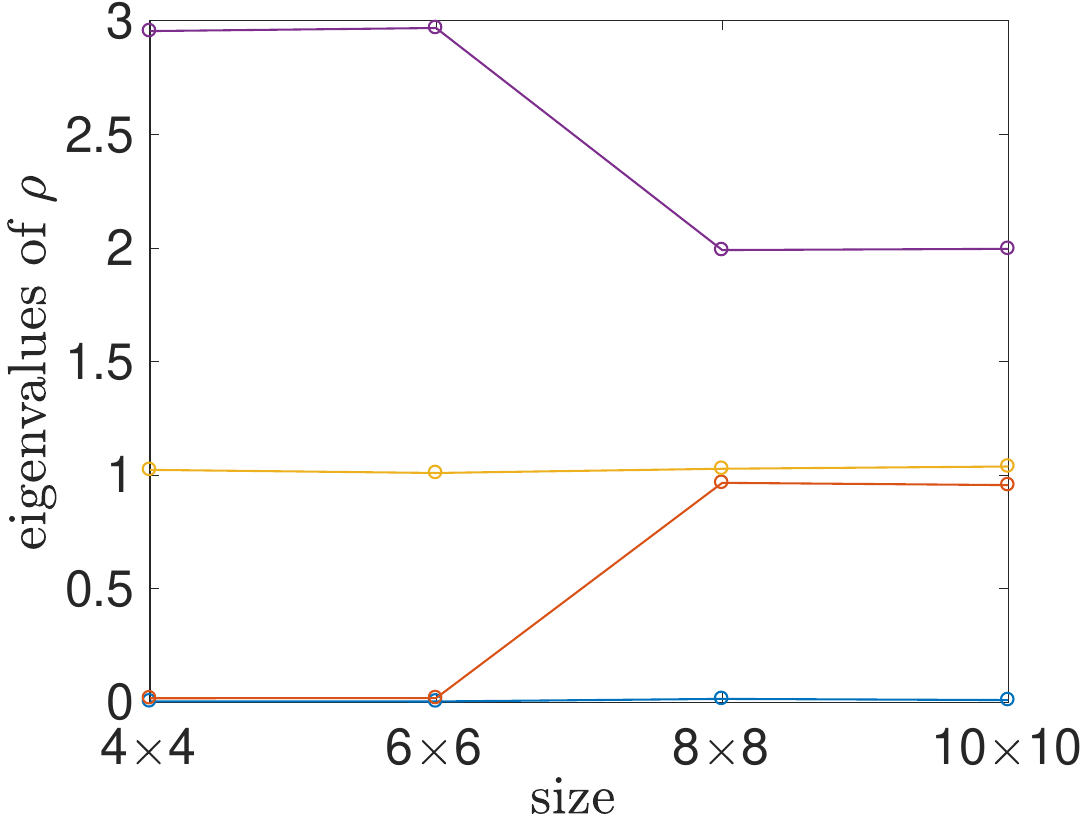}
}
\subfigure[GSD in the polarized phase]{
\includegraphics[width=7.0cm]{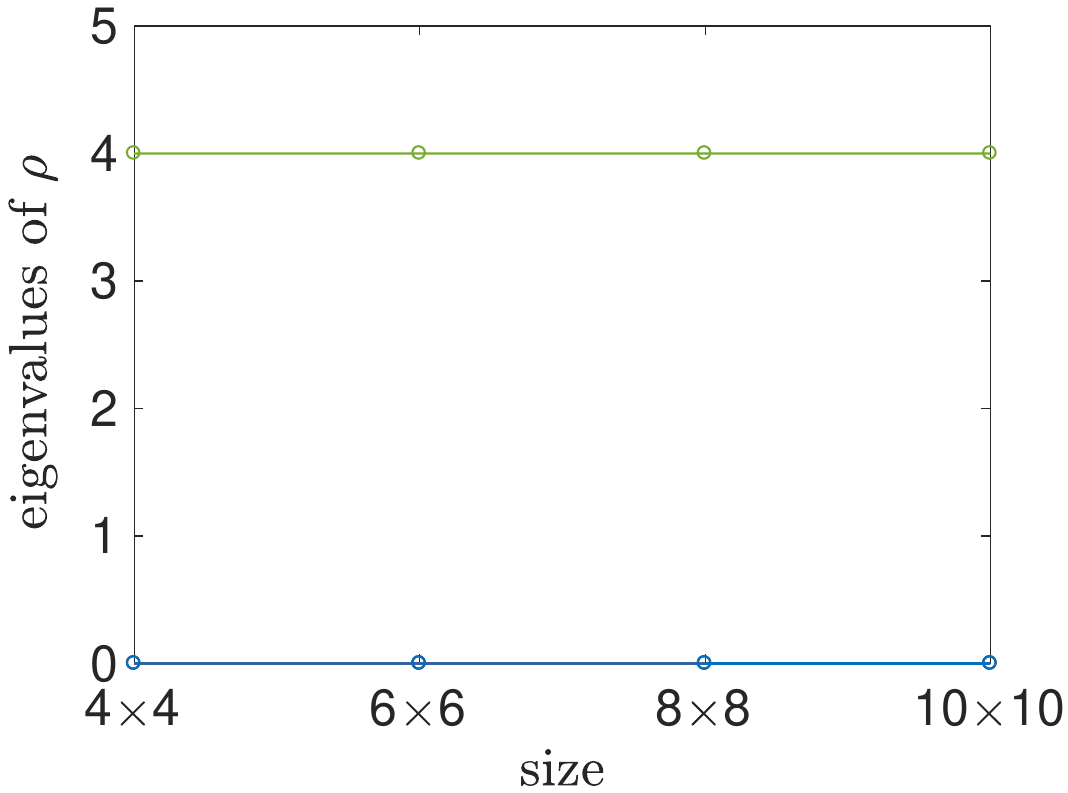}
}
\caption{\textbf{Eigenvalues of the fidelity matrix for the ground states on a torus.}
(A) In the QSL phase, all the eigenvalues are approaching 1 with the increasing size, indicating that in thermodynamic limit the GSD is 4 and the $Z_2$ gauge field is deconfined. (B) In the polarized phase, the GSD is one, indicating $Z_2$ confinement.}\label{fig:GSD}
\end{figure*}

Since the inserting of a $\pi$ flux in one of the holes is equivalent to change the boundary condition of the mean field Hamiltonian in the corresponding direction from the periodic one to anti-periodic one (vice versa).  If we note periodic and anti-periodic boundary conditions as + and $-$ respectively, then there are four different combinations of boundary conditions for the $x$- and $y$- directions, namely $(+,+), (+,-), (-,+), (-,-)$. The four boundary conditions results in four projected states, namely $|\psi_\alpha \rangle=P_G| \alpha \rangle_{\rm mf}$ where $|\alpha \rangle_{\rm mf}$ denotes the mean field ground state with boundary condition $\alpha$. If the ground state degeneracy(GSD) of the spin model on the torus is four, then $|\psi_{(+,+)}\rangle,  |\psi_{(+,-)}\rangle, |\psi_{(-,+)}\rangle, |\psi_{(-,-)}\rangle$ should be linearly independent. Otherwise, if the GSD is 1, then the above four states are essentially the same.

To judge the GSD, we calculate the fidelity matrix of the above four states with $\rho_{\alpha\beta}=\langle\psi_\alpha|\psi_\beta\rangle$. The four eigenvalues of the matrix $\rho$ reflects the GSD. If all of the eigenvalues are of order 1, as shown in Fig.~\ref{fig:GSD}A for the partially polarized QSL phase (the fact that last eigenvalue is quit small may owe to the small $Z_2$-flux excitation gap and the finite size effect), then the four states are linearly independent and consequently the GSD is four. The four-fold degeneracy of  the ground states on a torus indicates the deconfinement of the $Z_2$ gauge field. On the other hand, if one of the eigenvalue is close to 4 and the other three are close to 0, as shown in Fig.~\ref{fig:GSD}B for the polarized trivial phase, then the GSD is one which indicates the $Z_2$ confinement.




\subsubsection{Angular dependence of the magnetic field} 

Now we fix the strength of the external magnetic field $\pmb B$ and vary its direction. Then we study the off-diagonal magnetization of the ground state, namely the magnetization in the direction perpendicular to $\pmb B$. The off-diagonal magnetization, which can be read out from the variational parameters $\pmb M_\perp\propto  \tilde{\pmb B}_\perp=\tilde{\pmb B}\times \pmb B/|\pmb B|$,  is closely related to the experimentally measured torque $\pmb\tau=\mu_B\pmb M\times \pmb B=\mu_B\pmb M_\perp \times \pmb B$.

\begin{figure*}[htbp]
\includegraphics[width=8.0cm]{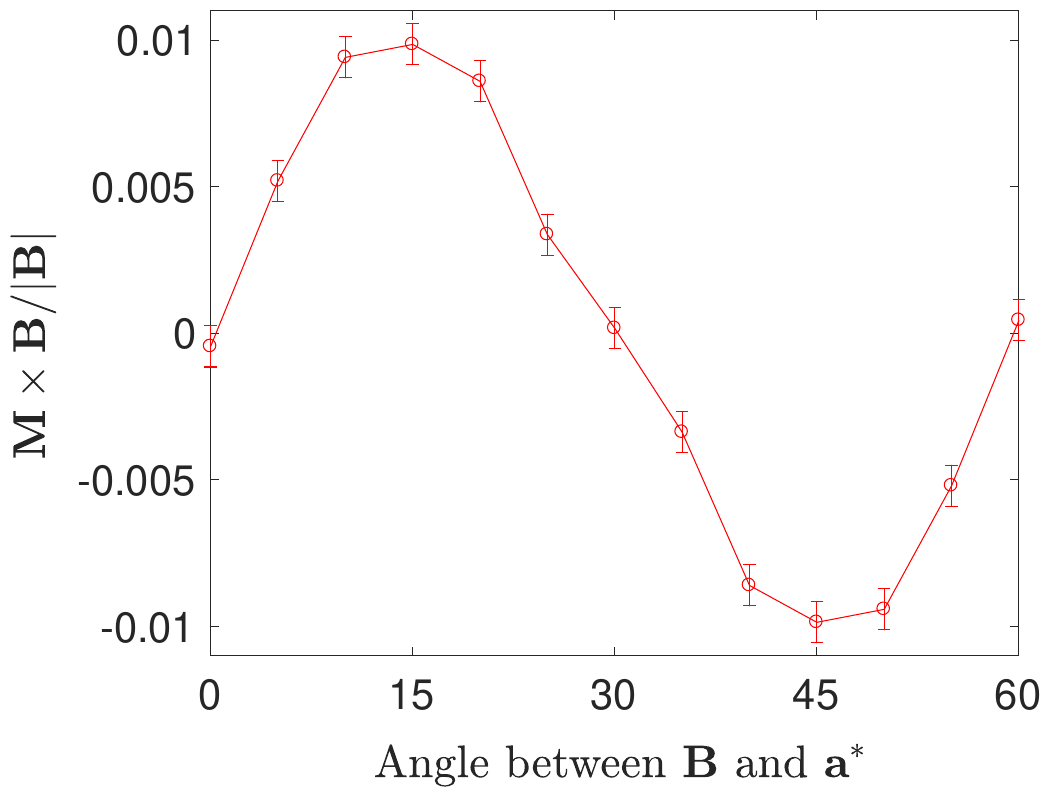}
\caption{ {\bf Calculated angle-dependent off-diagonal magnetization with fixed field strength.} The angle-dependence exhibits a 6-fold periodicity.}\label{fig:angle}
\end{figure*}

In the partially polarized QSL phase, there is no symmetry breaking. Therefore, when changing the angle between $\pmb B$ and $\bf a^*$, it is expected that the angle dependence of the off-diagonal magnetization should respect all the symmetries of the Honeycomb lattice. The numerical result is shown in Fig.~\ref{fig:angle}, which indicates a 6-fold periodicity and is consistent with the expectation.

\subsection{Excited states and Dynamical structure factor}\label{sec:DSF}

\subsubsection{Mean-field spectrum of the DSF}

From fluctuation-dissipation relation, the dynamical structure factor (DSF) $S^{mn}(\pmb k,\omega), m,n=x,y,z$ is defined by
\Beq
S^{mn}(\pmb k,\omega)= -{1\over\pi}[1+n_B(\omega)]{\chi''}^{mn}(\pmb k,\omega),
\Eeq
where $n_B(\omega)$ is the Bose-Einstein distribution and ${\chi''}^{mn}(\pmb k,\omega)=\lim_{\delta\to 0}{\rm Im\ } \chi^{mn}(\pmb k,\omega+i\delta)$ is the imaginary part of $\chi^{mn}(\pmb k,\omega)$ which is the analytic continuitioin of the finite temperature susceptibility $\chi^{mn}(\pmb k,i\omega)$. As an example, we illustrate the calculation of $\chi^{+-}(\pmb k,i\omega)$ in the following
{\small
\Beq
&&\chi^{+-}(\pmb k,i\omega) = \int_0^\beta  \langle T_\tau S^+_{\pmb k}(\tau) S^-_{-\pmb k}(0) \rangle e^{i\omega \tau} d\tau\\
&&\ \ \ = \int_0^\beta \sum_{\pmb q \pmb p,\alpha \beta} \langle T_\tau c^{\alpha \dag}_{\up,\pmb q}(\tau) c^\alpha_{\dn,\pmb q+\pmb k}(\tau) c^{\beta \dag}_{\dn,\pmb p+\pmb k}(0) c^\beta_{\up,\pmb p}(0)\rangle e^{i\omega \tau} d\tau 
\Eeq
}
where $S^\pm=S_x\pm iS_y$ and the superscript $\alpha,\beta$ sums over the  A,B sub-lattices, $T_\tau$ means time order and $\langle\rangle$ stands for thermal average.  If we abbreviate the spin, sub-lattice and particle-hole indices of $C$-fermions (here $C_k=(c_{\up A,\pmb k},c_{\dn A,\pmb k},c_{\up B,\pmb k},c_{\dn B,\pmb k},c^\dag_{\up A,-\pmb k},c^\dag_{\dn A,-\pmb k},c^\dag_{\up B,-\pmb k},c^\dag_{\dn B,-\pmb k})^T$) in $\chi^{+-}(\pmb k,i\omega)$ as $a,b,c,d$, namely, $\langle T_\tau c^{\alpha \dag}_{\up,\pmb q}(\tau) c^\alpha_{\dn,\pmb q+\pmb k}(\tau) c^{\beta \dag}_{\dn,\pmb p+\pmb k}(0) c^\beta_{\up,\pmb p}(0)\rangle =\langle T_\tau c^{a \dag}_{\pmb q}(\tau) c^b_{\pmb q+\pmb k}(\tau) c^{c \dag}_{\pmb p+\pmb k}(0) c^d_{\pmb p}(0)\rangle $, using Wick theorem, we have,
{\small
\Beq
&&\chi^{+-}(\pmb k,i\omega)={1\over V}\int_0^\beta \sum_{ \pmb q \pmb p,\{abcd\}} \langle T_\tau c^{a \dag}_{\pmb q}(\tau) c^b_{\pmb q+\pmb k}(\tau) c^{c \dag}_{\pmb p+\pmb k}(0) c^d_{\pmb p}(0)\rangle e^{i\omega\tau}d\tau\\
&&\ \ \ ={1\over V}\int_0^\beta \sum_{ \pmb q \pmb p,\{abcd\}}
\left[ -\langle T_\tau  c^d_{\pmb p}(0) c^{a \dag}_{\pmb q}(\tau) \rangle \langle T_\tau c^b_{\pmb q+\pmb k}(\tau) c^{c \dag}_{\pmb p+\pmb k}(0) \rangle \right.\\
&&\ \ \ \ \ \ +\left. \langle T_\tau c^{c+4}_{\pmb p+\pmb k}(0) c^{a \dag}_{\pmb q}(\tau) \rangle \langle T_\tau c^b_{\pmb q+\pmb k}(\tau) c^{d+4,\dag}_{\pmb p}(0)\rangle\right] e^{i\omega \tau}  d\tau\\
&&\ \ \ ={1\over V}\int_0^\beta \sum_{ \pmb q, \{abcd\}}
\left[ -\langle T_\tau  c^d_{\pmb q}(0) c^{a \dag}_{\pmb q}(\tau) \rangle \langle T_\tau c^b_{\pmb q+\pmb k}(\tau) c^{c \dag}_{\pmb q+\pmb k}(0) \rangle \right.\\
&&\ \ \ \ \ \ +\left. \langle T_\tau c^{c+4}_{-\pmb q}(0) c^{a \dag}_{\pmb q}(\tau) \rangle \langle T_\tau c^b_{\pmb q+\pmb k}(\tau) c^{d+4, \dag}_{-(\pmb q+\pmb k)}(0)\rangle\right]e^{i\omega \tau} d\tau
\Eeq
}

At mean field lever, the $C$ fermions are diagonalized into the Bogoliubov particles, and the dynamic structure factors can be evaluated numerically.  The complete dynamic structure factor $S(\pmb k,\omega) ={1\over2}(S^{+-}(\pmb k,\omega)+S^{-+}(\pmb k,\omega))+S^{zz}(\pmb k,\omega)$ for a system with $30\times 30\times 2=1800$ sites is shown in Fig.\ref{INS}A, where the spinon continuum can be clearly seen in the range $\omega<3$meV.



\begin{figure*}[t]
\subfigure[ $S(\pmb k,\omega)$ from Mean field]{\label{fig:DSFmf}
\centering
\includegraphics[width=7.6cm]{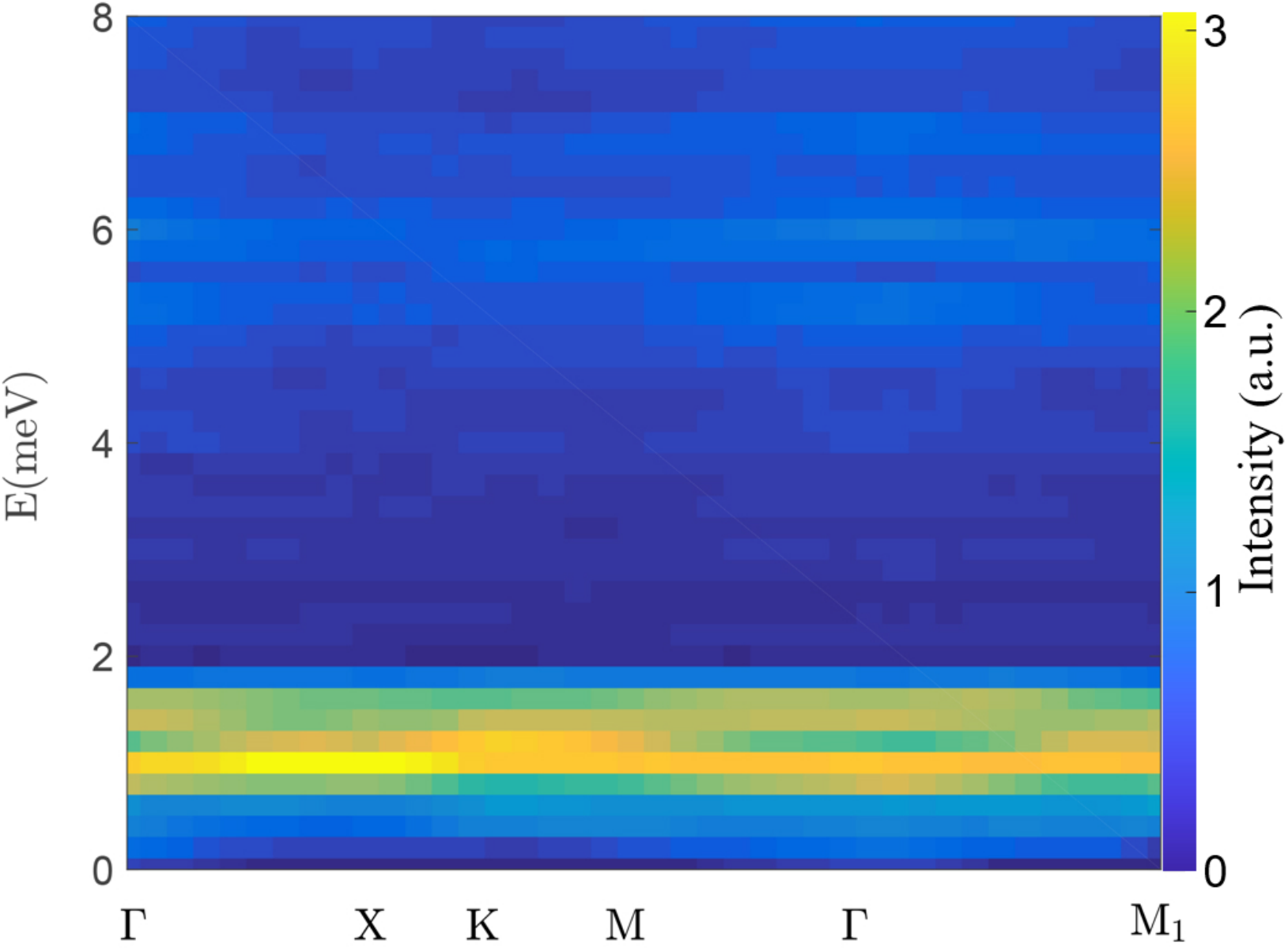}}
\subfigure[ $S^{zz}(\pmb k,\omega)$ from VMC]{\label{fig:DSFvmc}
\centering
\includegraphics[width=7cm]{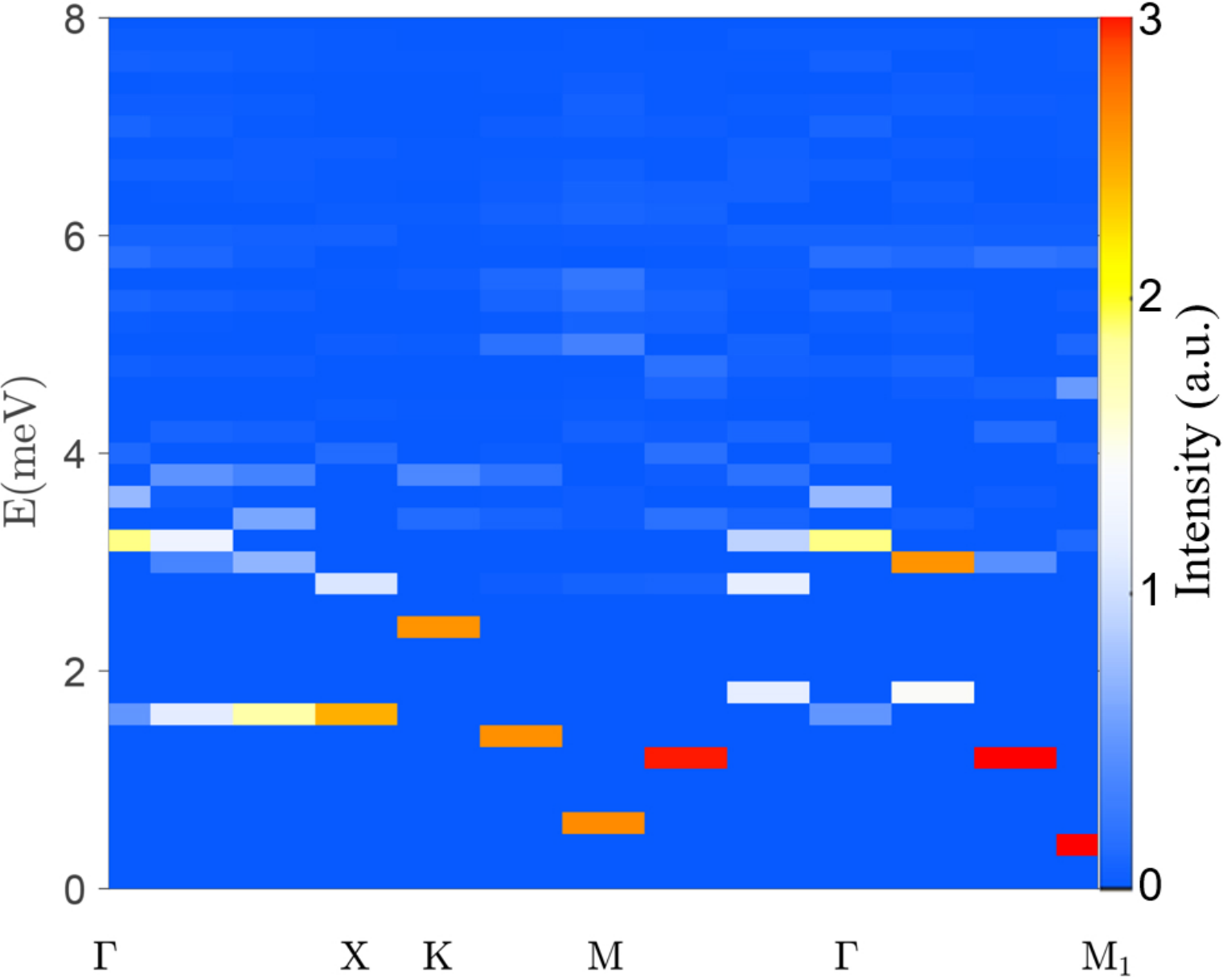}
}
\caption{\textbf{Calculated dynamic structure factors.} The dynamic structure factors calculated from (A) mean field states with $30\times 30\times2$ sites and (B) Gutzwiller projected states with $6\times6\times 2$ sites, respectively. }\label{INS}
\end{figure*}

\subsubsection{VMC results of the DSF }

Now we try to construct the low-energy excitations via Gutzwiller projected spinon excited states
\beq\label{spinonpq}
|\pmb k;({\pmb k - \pmb q})_m,({\pmb q})_n\rangle = P_G f_{\pmb k-\pmb q,m}^\dag f_{\pmb q,n}^{\dag} |\Psi_{\rm mf} \rangle,
\eeq
where $f_{\pmb k-\pmb q,m}$ and $f_{\pmb q,n}$ are eigen particles of the mean-field Hamiltonian. Here $m$ and $n$ are the band indices, $\pmb k$ and $\pmb q$ are the lattice momentum according to the translation operators.

The above two-spinon excitations form a continuum and do not correctly describe the low-energy excitations of the system. The failure of the state (\ref{spinonpq}) in describing  the low-energy excitations is owing to the strong gauge interactions between the spinons which are not correctly addressed. To partially solve this problem, we diagonalize the original Hamiltonian in the subspace spanned by the two-spinon excitation continuum. Specifically, we calculate the matrix elements of the matrix $\mathscr H(\pmb k)$ with
\Beq
\mathscr H_{pmn,qm'n'}(\pmb k) = \langle  \pmb k; (\pmb k-\pmb p)_m, (\pmb p)_n |H| \pmb k; (\pmb k-\pmb q)_{m'}, (\pmb q)_{n'} \rangle,
\Eeq
where $\pmb k$ is the total momentum of the state with a pair of spinon excitations. 

Since the states in the two-spinon continuum are not orthogonal, we need to calculate the metric matrix $g$ formed by the overlap of the states,
\[
g_{pmn,qm'n'}(\pmb k) =\langle  \pmb k; (\pmb k-\pmb p)_m, (\pmb p)_n | \pmb k; (\pmb k-\pmb q)_{m'}, (\pmb q)_{n'} \rangle.
\]
Hence, the eigenproblem of $\mathscr H(\pmb k)$ should be calculated by
\[
 g^{-1}(\pmb k)\mathscr H(\pmb k) U = U\cdot {\rm diag}(\epsilon_1, \epsilon_2, ... ), 
\]
where the eigenvalues $\epsilon_1,\epsilon_{2}, ... $ are the `renormalized' energy of the excitations. 

The renormalized eigenfunction $|\pmb k\rangle_{\rm rn}$ is given by
\Beq
|\pmb k\rangle_{\rm rn} = \sum_{\pmb q\in BZ} \mathscr F(qmn)|\pmb k; (\pmb k-\pmb q)_m,(\pmb q)_n\rangle,
\Eeq
where $\mathscr F(qmn)$ is the eigenvector of the matrix $\mathscr H(\pmb k)$.

The more reliable DSF can be calculated using the renormalized energy and eigenstates. For instance, the $zz$-component of the DSF is expressed as
\Beq
S^{zz}(\pmb q, \omega) = \sum_{n}\left|\left\langle\Psi_{n}^{\pmb q}|S_{\pmb q}^{z}|\Psi_{0}\right\rangle\right|^{2} \delta\left(\omega-E_{n}^{\pmb q}+E_{0}\right)
\Eeq
where $\left|\Psi_{0}\right\rangle$ is the variational ground state with energy $E_{0}$ and $\left|\Psi_{n}^{\pmb q}\right\rangle$ is the n-th `renormalized' two-spinon excited state with momentum $\pmb q$ and energy $E_{n}^{\pmb q}$. We note that $S_{\pmb q}^{z}=\frac{1}{\sqrt{L}} \sum_{\pmb r} \exp [i \pmb q \cdot \pmb r] S_{\pmb r}^{z}$ is the Fourier-transformed spin operator for the component $z$, with $L$ being the number of sites in the system.


However, due to the `renormalizing' process, the computational complexity increases dramatically with system size. The data shown in Fig.\ref{INS}B is the result for a system with size $6\times 6\times2=72$ sites. In Fig.\ref{INS}B, the magnon modes as bound states of spinons can be clearly seen. However, due to finite size effect, the excitation gap of the magnons and the energy of the spinon continuum are exaggerated.

Noticing that the spinons are deconfined in the low-energy limit, the mean field results are qualitatively correct (but the bound states cannot be obtained at mean field level).  Therefore, in the main text, we have combined the mean field result (with the spinon continuum) and the VMC result (with the magnon modes) to estimate the dynamic structure factor at large size limit.

\section{Linear spin wave}\label{sec:SW}

\subsection{\bf Ordered phase}
In the zigzag phase, the magnetic unit cell contains 4 sites, as shown in Fig.\ref{fig:LSW}A. To obtain the linear spin wave (LSW) dispersion, we firstly adopt a new spin frame such that the new $z$-axis are parallel to the zigzag order, namely
\begin{equation}
\begin{aligned}
\begin{pmatrix}S_i^{\prime x} \\ S_i^{\prime y} \\ S_i^{\prime z}\end{pmatrix} = \mathcal R_\alpha \begin{pmatrix}S_i^{x} \\ S_i^{y} \\ S_i^{z}\end{pmatrix}
\end{aligned}
\end{equation}
where $\mathcal R_\alpha$ is a $SO(3)$ matrix with $\alpha=$R,B stands for the red/blue sublattice index. Notice that the new $z$-axes in the red sublatice and the blue sublattice are opposite to each other. Thus the original $K$-$J_1$-$\Gamma$-$\Gamma'$-$J_3$ model is transformed into a new one under the above rotations,
\beq\label{rotH}
H = \sum_{\langle i,j \rangle }\pmb S^{\prime\mathrm{T}}_i{H}^{'ij} \pmb S^{\prime}_j - g\mu_B\sum_{i}\pmb B_i^{\prime}\cdot\pmb{S}^{\prime}_i,
\eeq
where ${H}^{'ij}=\mathcal{R}_i{H}^{ij}\mathcal{R}_j^\mathrm{T}$ (with ${H}^{ij}$ the original interaction matrix on the $ij$-bond), $\pmb B'_i= \mathcal{R}_i \pmb B$ with $\mathcal R_i = \mathcal R_R$ if $i$ belong to the red sublattice and $\mathcal R_i = \mathcal R_B$ for the blue sublattice, and $g=4.13$.\upcite{5}

In the new spin frame, we adopt the Holstein–Primakoff transformation $S_i^{\prime x} \sim{1\over2}(a_i^\dag+a_i), S_i^{\prime y} \sim{i\over2}(a_i^\dag-a_i), S_i^{\prime z} ={1\over2}-a_i^{\dag}a_i$ on the A-sublattice and $S_i^{' x} \sim{1\over2}(b_i^\dag+b_i), S_i^{\prime y} \sim{i\over2}(b_i^\dag-b_i), S_i^{\prime z} ={1\over2}-b_i^{\dag}b_i$ on the B-sublattice,
where the bosons satisfy the usual commutation relations $[a_i, a_j^\dagger] = \delta_{ij}, [b_i, b_j^\dagger] = \delta_{ij}$  and $[a_i, b_j^\dagger] = 0$.
Substituting the above formulas into the rotated Hamiltonian (\ref{rotH}) and keeping the quadratic terms, we obtain the following Hamiltonian on the Fourier bases,
\beq\label{HSW}
H_{\rm SW} = \sum_{\pmb k} \Psi^\dagger(\pmb k) \mathcal{H}(\pmb k) \Psi(\pmb k),
\eeq
where $\Psi^\dagger(\pmb k)=(a_{R,\pmb k}^{\dag}, b_{B,\pmb k}^{\dag}, a_{B,\pmb k}^{\dag}, b_{R,\pmb k}^{\dag}, a_{R,\pmb {-k}}, b_{B,\pmb {-k}}, a_{B,\pmb {-k}}, b_{R,\pmb {-k}})$  with $a_{R/B,\pmb k}^\dag$ ($b_{R/B,\pmb k}^\dag$) being the magnon creation operator on A (B) and red/blue sublattices.
Diagonalizing the above Hamiltonian using the Bosonic Bogoliubov transformation, we obtain the magnon excitations spectrum at zero field, as shown in Fig.\ref{fig:LSW}B.

\subsection{\bf Disordered phases}
In the disordered phases, the spins are (partially) polarized due to the magnetic field. One can also calculate the spin wave spectrum assuming that the fully polarized state is the classical ground state. To this end, we perform an uniform orthogonal transformation to rotate the new $z$-axis to the field direction ($\bf a^*)$, and
\begin{equation}
\begin{aligned}
\begin{pmatrix}S_i^{\prime x} \\ S_i^{\prime y} \\ S_i^{\prime z}\end{pmatrix} =
\begin{pmatrix}\frac{1}{\sqrt{6}} & \frac{1}{\sqrt{6}} & \frac{-2}{\sqrt{6}} \\ \frac{1}{\sqrt{3}} & \frac{1}{\sqrt{3}} & \frac{1}{\sqrt{3}}\\\frac{1}{\sqrt{2}} & \frac{-1}{\sqrt{2}} & 0  \end{pmatrix} \begin{pmatrix}S_i^{x} \\ S_i^{y} \\ S_i^{z}\end{pmatrix},
\end{aligned}
\end{equation}
and then introduce the Holstein–Primakoff transformation to obtain the spin wave Hamiltonian and the corresponding spectrum.

\begin{figure*}[t]
\includegraphics[width=3cm]{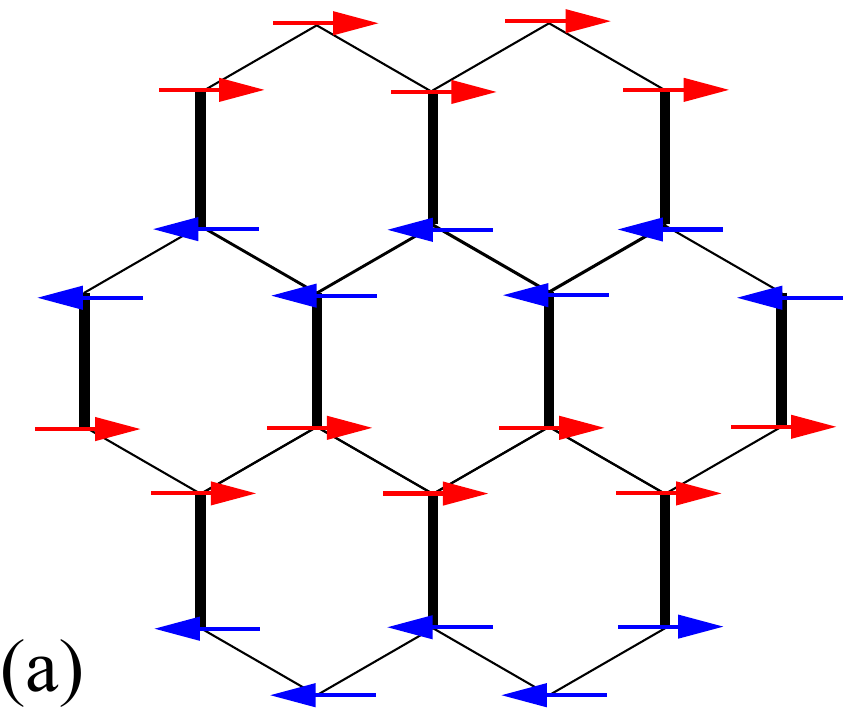}\! \! \!
\includegraphics[width=4cm]{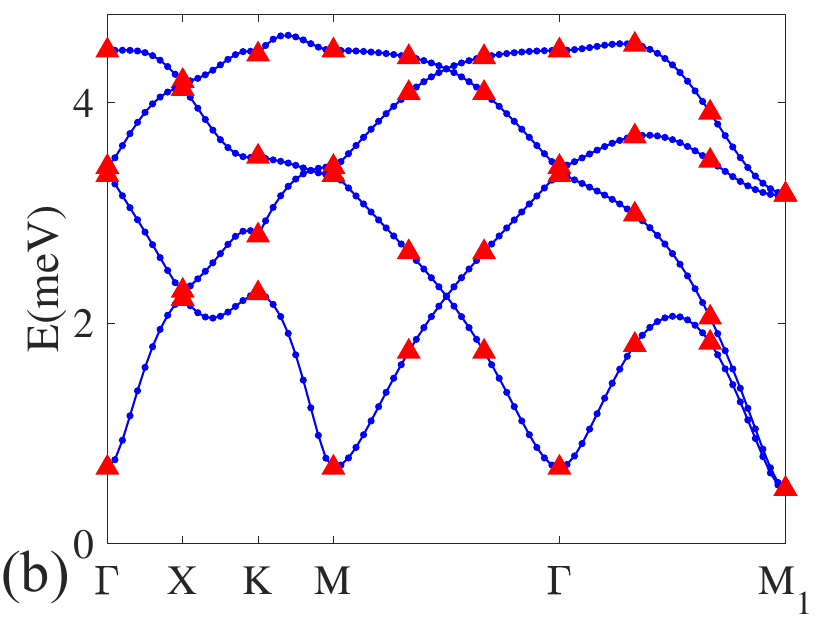}\! \! \!
\includegraphics[width=4cm]{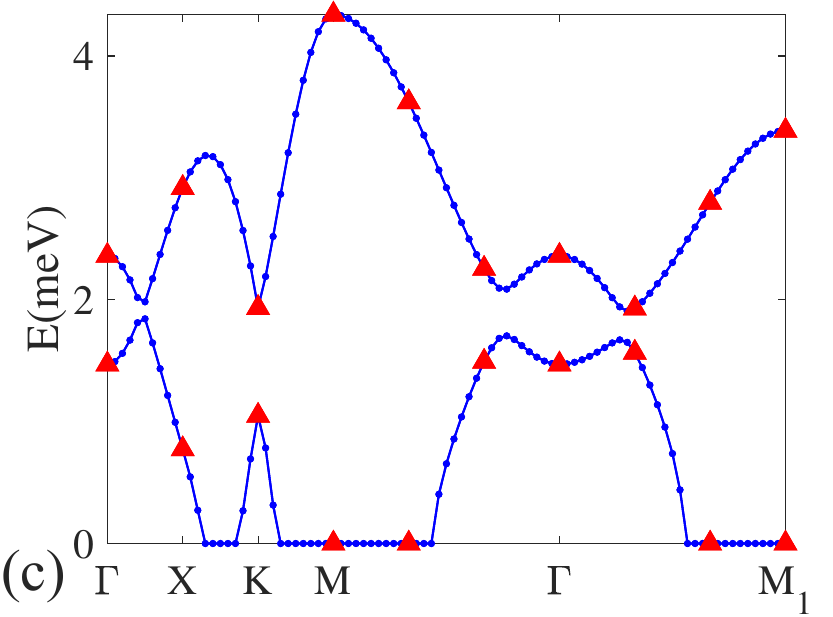}\! \! \!
\includegraphics[width=4cm]{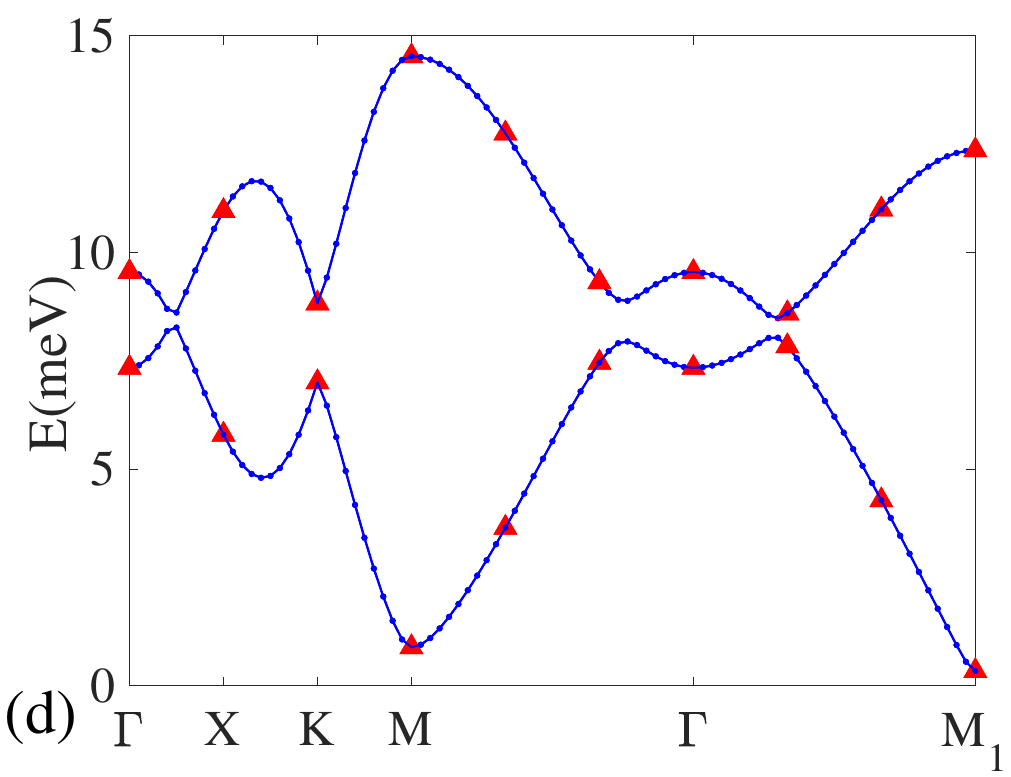}
\caption{\textbf{Linear spin wave dispersion.} (A) The Zigzag order on honeycomb lattice. (B) Linear spin wave of the Zigzag phase at zero field. (C) Linear spin wave of the disordered phase at $\mu_B B/|K|=0.56$. (D) Linear spin wave of the polarized phase at $\mu_BB/|K|=0.95$.}\label{fig:LSW} 
\end{figure*}

In the partially polarized QSL region (for instance at $\mu_B B/|K|$=0.56),  the LSW based on the fully polarized state is unstable, namely, the magnon spectrum has nonzero imaginary part when the real part goes to zero, as shown in Fig.\ref{fig:LSW}(C). The instability of the spin wave indicates that the polarized state is not the ground state in the intermediate field region. In contrast, when the field is strong enough (for instance at $\mu_BB/|K|$=0.95), the LSW based on the polarized state is fully gapped and stable [see Fig.\ref{fig:LSW}D].


\section{Random phase approximations}\label{sec:RY}

\begin{figure*}[t]
	\subfigure[Dynamical structure factor obtained by RPA]{\label{fig:DSF}
		\centering
		\includegraphics[width=8.63cm]{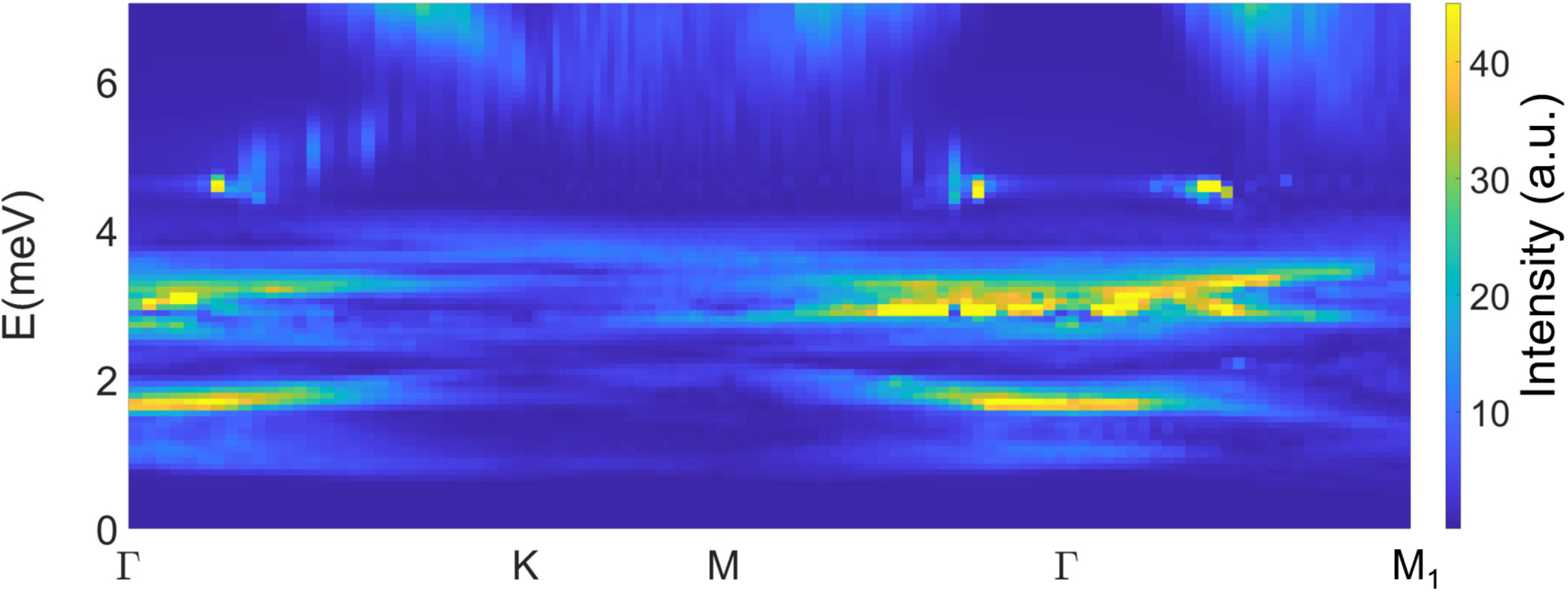}}
	\caption{\textbf{Calculated dynamic structure factors.} The dynamic structure factors obtained by RPA with $30\times 30\times2$ sites at $U=6$. }
\end{figure*}

\begin{figure*}[t]
	\subfigure[ $S^{+-}(\pmb k,\omega)$]{\label{fig:Transverse_DSF}
		\centering
		\includegraphics[width=7.63cm]{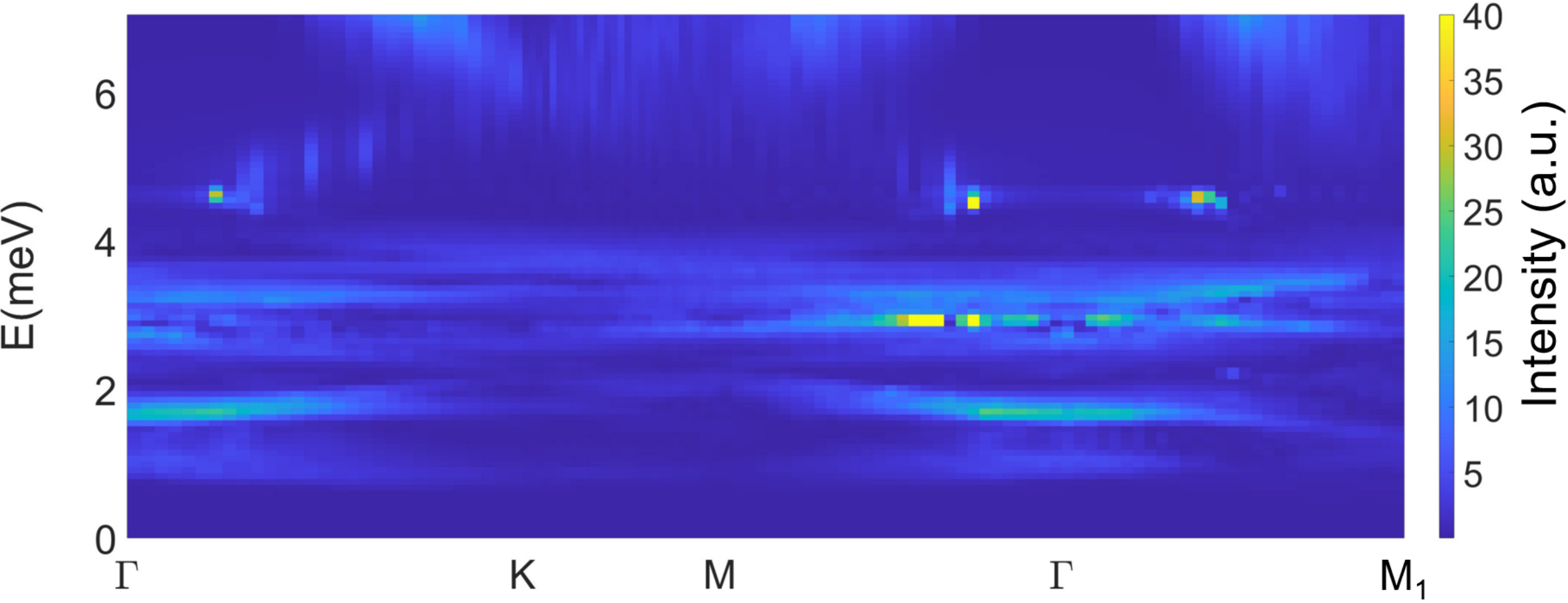}}
	\subfigure[ $S^{zz}(\pmb k,\omega)$]{\label{fig:Longitude_DSF}
		\centering
		\includegraphics[width=7.63cm]{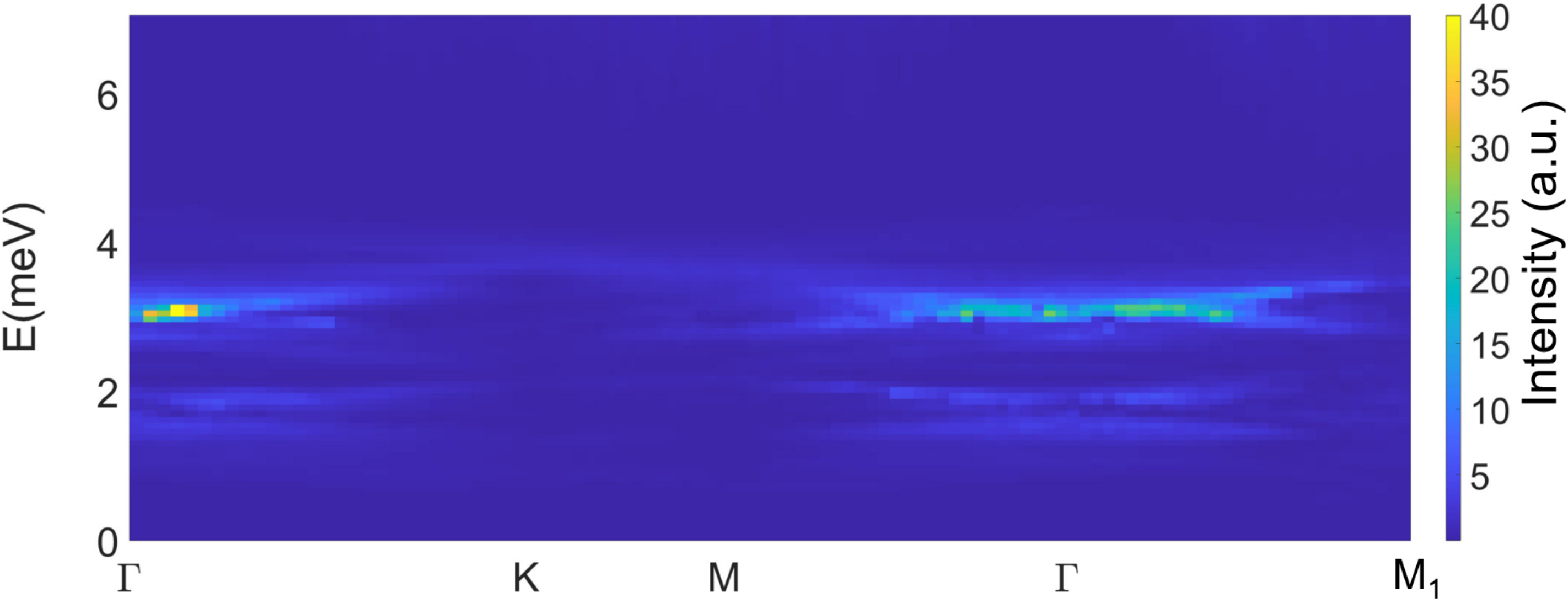}
	}
	\caption{\textbf{The transverse and longitude dynamical structure factors.} (a) The transverse dynamic structure factors obtained by RPA with $30\times 30\times2$ sites at $U=6$  and (b) The longitude dynamic structure factors, respectively. }
\end{figure*}

Using the determined model parameters, we extend our analysis to include spin fluctuations beyond the mean-field level by employing the random phase approximation (RPA).\upcite{RongYu, TLi} The RPA provides a means to incorporate the dynamical behavior of the interacting system by applying corrections to the bare susceptibility $\chi(0)$. The expression for the RPA-corrected susceptibility is given by
\begin{equation}
	\chi(q,i\omega_n)=[I+V(q)\chi^{(0)}(q,i\omega_n)]^{-1}\chi^{(0)}(q,i\omega_n)
\end{equation}
where $I$ denotes the identity matrix and the kernel V(q) is defined as:
\begin{equation}
	V(q) = J(q)-U
\end{equation}
In this equation, $U$ is the strength of a phenomenological Hubbard term introduced for penalty of double occupancy of spinons, and $J(q)$ denotes the Fourier transformation of the exchange interactions.
At zero temperature, the bare spin susceptibility $\chi(0)$ is given by:
\begin{equation}
	\chi^{(0)i,j}_{\mu,\nu}(q,\omega) =\frac{1}{4} \sum_{k,n}[G^{\nu\mu}(k+q,\omega_n+\omega)G^{\mu\nu}(k,\omega_n) -\bar{F}^{\nu\mu}(k+q,\omega_n+\omega)F^{\mu\nu}(k,\omega_n)].
\end{equation}
Here, indices $\mu$ and $\nu$ refer to the A and B sublattices, respectively, and $i$, $j = x$, $y$, $z$ represent indices of spin components. The quantities $G$ and $F$ correspond to the normal and anomalous parts of the spinon Green's function, respectively. The complete dynamical structure factor is $\chi(k.\omega)=\chi^{+-}(k,\omega)+\chi^{-+}(k,\omega)+\chi^{zz}(k,\omega)$ shown in Fig.\ref{fig:DSF}. Additionally, we conduct calculations for the transverse dynamical susceptibility and the longitudinal dynamical susceptibility, as depicted in Fig.\ref{fig:Transverse_DSF} and Fig.\ref{fig:Longitude_DSF}, respectively.
Dispersive resonance modes are clearly seen at $E\approx2$ meV and $E\approx3$ meV. These results agree well with both experimental data and VMC results. Our results also reveal a dispersive resonance mode centered at $\Gamma$ point between $4$ and $6$ meV, a feature also captured by experimental data, but with strong damping.



\end{document}